\documentclass{article}

\usepackage{graphicx} 
\usepackage[margin=2.2cm]{geometry}
\usepackage{subcaption}
\usepackage{amsmath}
\usepackage{titling}
\usepackage{float}
\usepackage{hyperref}
\usepackage[
  backend=biber,
  style=numeric-comp, 
    sortcites=true, 
  sorting=none,    
  maxbibnames=10,  
  minbibnames=10    
]{biblatex}
\usepackage{authblk}

\title{Automated Compilation Including Dropouts:
\vspace{0.2cm}\\ \Large{Tolerating Defective Components in Stabiliser Codes}}
\author{Stasiu Wolanski \thanks{\texttt{stasiu.wolanski@riverlane.com}}\\
\textit{Riverlane, St Andrew's House, Cambridge, UK} \vspace{0.2cm}\\
\textit{
Dept. of Physics and Astronomy, University College London, UK}}
\date{November 2025}

\newcommand{\subfiglab}[1]{\textit{\textbf{#1}}}
\begin{document}
\newcommand{\bigurl}{https://stasiu51.github.io/Shatter/##circuit=####!_SHEET_NAME=QUBITS;####!_SHEET_NAME=E;####!_SHEET_NAME=UNTX;####!_SHEET_NAME=UNTZ;####!_SHEET_NAME=ANTIX;####!_SHEET_NAME=ANTIZ;####!_SHEET_NAME=PRODX;####!_SHEET_NAME=PRODZ;####!_SHEET_NAME=GAUGEX;####!_SHEET_NAME=GAUGEZ;####!_EMBEDDING_TYPE=PLANE;####!_QUBIT_Q=0_SHEET=QUBITS_COLOUR=gold;Q(0,0)_0;####!_QUBIT_Q=1_SHEET=QUBITS_COLOUR=gold;Q(0,2)_1;####!_QUBIT_Q=2_SHEET=QUBITS_COLOUR=gold;Q(0,4)_2;####!_QUBIT_Q=3_SHEET=QUBITS_COLOUR=gold;Q(0,6)_3;####!_QUBIT_Q=4_SHEET=QUBITS_COLOUR=gold;Q(0,8)_4;####!_QUBIT_Q=5_SHEET=QUBITS_COLOUR=gold;Q(0,10)_5;####!_QUBIT_Q=6_SHEET=QUBITS_COLOUR=gold;Q(0,12)_6;####!_QUBIT_Q=7_SHEET=QUBITS_COLOUR=gold;Q(0,14)_7;####!_QUBIT_Q=8_SHEET=QUBITS_COLOUR=gold;Q(0,16)_8;####!_QUBIT_Q=9_SHEET=QUBITS_COLOUR=gold;Q(0,18)_9;####!_QUBIT_Q=10_SHEET=QUBITS_COLOUR=gold;Q(0,20)_10;####!_QUBIT_Q=11_SHEET=QUBITS_COLOUR=gold;Q(1,1)_11;####!_QUBIT_Q=12_SHEET=QUBITS_COLOUR=gold;Q(1,3)_12;####!_QUBIT_Q=13_SHEET=QUBITS_COLOUR=gold;Q(1,5)_13;####!_QUBIT_Q=14_SHEET=QUBITS_COLOUR=gold;Q(1,7)_14;####!_QUBIT_Q=15_SHEET=QUBITS_COLOUR=gold;Q(1,9)_15;####!_QUBIT_Q=16_SHEET=QUBITS_COLOUR=gold;Q(1,11)_16;####!_QUBIT_Q=17_SHEET=QUBITS_COLOUR=gold;Q(1,13)_17;####!_QUBIT_Q=18_SHEET=QUBITS_COLOUR=gold;Q(1,15)_18;####!_QUBIT_Q=19_SHEET=QUBITS_COLOUR=gold;Q(1,17)_19;####!_QUBIT_Q=20_SHEET=QUBITS_COLOUR=gold;Q(1,19)_20;####!_QUBIT_Q=21_SHEET=QUBITS_COLOUR=gold;Q(1,21)_21;####!_QUBIT_Q=22_SHEET=QUBITS_COLOUR=gold;Q(2,0)_22;####!_QUBIT_Q=23_SHEET=QUBITS_COLOUR=gold;Q(2,2)_23;####!_QUBIT_Q=24_SHEET=QUBITS_COLOUR=gold;Q(2,4)_24;####!_QUBIT_Q=25_SHEET=QUBITS_COLOUR=gold;Q(2,6)_25;####!_QUBIT_Q=26_SHEET=QUBITS_COLOUR=gold;Q(2,8)_26;####!_QUBIT_Q=27_SHEET=QUBITS_COLOUR=gold;Q(2,10)_27;####!_QUBIT_Q=28_SHEET=QUBITS_COLOUR=gold;Q(2,12)_28;####!_QUBIT_Q=29_SHEET=QUBITS_COLOUR=gold;Q(2,14)_29;####!_QUBIT_Q=30_SHEET=QUBITS_COLOUR=gold;Q(2,16)_30;####!_QUBIT_Q=31_SHEET=QUBITS_COLOUR=gold;Q(2,18)_31;####!_QUBIT_Q=32_SHEET=QUBITS_COLOUR=gold;Q(2,20)_32;####!_QUBIT_Q=33_SHEET=QUBITS_COLOUR=gold;Q(3,1)_33;####!_QUBIT_Q=34_SHEET=QUBITS_COLOUR=gold;Q(3,3)_34;####!_QUBIT_Q=35_SHEET=QUBITS_COLOUR=gold;Q(3,5)_35;####!_QUBIT_Q=36_SHEET=QUBITS_COLOUR=gold;Q(3,7)_36;####!_QUBIT_Q=37_SHEET=QUBITS_COLOUR=gold;Q(3,9)_37;####!_QUBIT_Q=38_SHEET=QUBITS_COLOUR=gold;Q(3,11)_38;####!_QUBIT_Q=39_SHEET=QUBITS_COLOUR=gold;Q(3,13)_39;####!_QUBIT_Q=40_SHEET=QUBITS_COLOUR=gold;Q(3,15)_40;####!_QUBIT_Q=41_SHEET=QUBITS_COLOUR=gold;Q(3,17)_41;####!_QUBIT_Q=42_SHEET=QUBITS_COLOUR=gold;Q(3,19)_42;####!_QUBIT_Q=43_SHEET=QUBITS_COLOUR=gold;Q(3,21)_43;####!_QUBIT_Q=44_SHEET=QUBITS_COLOUR=gold;Q(4,0)_44;####!_QUBIT_Q=45_SHEET=QUBITS_COLOUR=gold;Q(4,2)_45;####!_QUBIT_Q=46_SHEET=QUBITS_COLOUR=gold;Q(4,4)_46;####!_QUBIT_Q=47_SHEET=QUBITS_COLOUR=gold;Q(4,6)_47;####!_QUBIT_Q=48_SHEET=QUBITS_COLOUR=gold;Q(4,8)_48;####!_QUBIT_Q=49_SHEET=QUBITS_COLOUR=gold;Q(4,10)_49;####!_QUBIT_Q=50_SHEET=QUBITS_COLOUR=gold;Q(4,12)_50;####!_QUBIT_Q=51_SHEET=QUBITS_COLOUR=gold;Q(4,14)_51;####!_QUBIT_Q=52_SHEET=QUBITS_COLOUR=gold_DEFECTIVE=true;Q(4,16)_52;####!_QUBIT_Q=53_SHEET=QUBITS_COLOUR=gold;Q(4,18)_53;####!_QUBIT_Q=54_SHEET=QUBITS_COLOUR=gold;Q(4,20)_54;####!_QUBIT_Q=55_SHEET=QUBITS_COLOUR=gold;Q(5,1)_55;####!_QUBIT_Q=56_SHEET=QUBITS_COLOUR=gold;Q(5,3)_56;####!_QUBIT_Q=57_SHEET=QUBITS_COLOUR=gold;Q(5,5)_57;####!_QUBIT_Q=58_SHEET=QUBITS_COLOUR=gold;Q(5,7)_58;####!_QUBIT_Q=59_SHEET=QUBITS_COLOUR=gold;Q(5,9)_59;####!_QUBIT_Q=60_SHEET=QUBITS_COLOUR=gold;Q(5,11)_60;####!_QUBIT_Q=61_SHEET=QUBITS_COLOUR=gold;Q(5,13)_61;####!_QUBIT_Q=62_SHEET=QUBITS_COLOUR=gold;Q(5,15)_62;####!_QUBIT_Q=63_SHEET=QUBITS_COLOUR=gold;Q(5,17)_63;####!_QUBIT_Q=64_SHEET=QUBITS_COLOUR=gold;Q(5,19)_64;####!_QUBIT_Q=65_SHEET=QUBITS_COLOUR=gold;Q(5,21)_65;####!_QUBIT_Q=66_SHEET=QUBITS_COLOUR=gold;Q(6,0)_66;####!_QUBIT_Q=67_SHEET=QUBITS_COLOUR=gold;Q(6,2)_67;####!_QUBIT_Q=68_SHEET=QUBITS_COLOUR=gold;Q(6,4)_68;####!_QUBIT_Q=69_SHEET=QUBITS_COLOUR=gold;Q(6,6)_69;####!_QUBIT_Q=70_SHEET=QUBITS_COLOUR=gold;Q(6,8)_70;####!_QUBIT_Q=71_SHEET=QUBITS_COLOUR=gold;Q(6,10)_71;####!_QUBIT_Q=72_SHEET=QUBITS_COLOUR=gold;Q(6,12)_72;####!_QUBIT_Q=73_SHEET=QUBITS_COLOUR=gold;Q(6,14)_73;####!_QUBIT_Q=74_SHEET=QUBITS_COLOUR=gold;Q(6,16)_74;####!_QUBIT_Q=75_SHEET=QUBITS_COLOUR=gold;Q(6,18)_75;####!_QUBIT_Q=76_SHEET=QUBITS_COLOUR=gold;Q(6,20)_76;####!_QUBIT_Q=77_SHEET=QUBITS_COLOUR=gold;Q(7,1)_77;####!_QUBIT_Q=78_SHEET=QUBITS_COLOUR=gold;Q(7,3)_78;####!_QUBIT_Q=79_SHEET=QUBITS_COLOUR=gold;Q(7,5)_79;####!_QUBIT_Q=80_SHEET=QUBITS_COLOUR=gold;Q(7,7)_80;####!_QUBIT_Q=81_SHEET=QUBITS_COLOUR=gold;Q(7,9)_81;####!_QUBIT_Q=82_SHEET=QUBITS_COLOUR=gold;Q(7,11)_82;####!_QUBIT_Q=83_SHEET=QUBITS_COLOUR=gold;Q(7,13)_83;####!_QUBIT_Q=84_SHEET=QUBITS_COLOUR=gold;Q(7,15)_84;####!_QUBIT_Q=85_SHEET=QUBITS_COLOUR=gold;Q(7,17)_85;####!_QUBIT_Q=86_SHEET=QUBITS_COLOUR=gold;Q(7,19)_86;####!_QUBIT_Q=87_SHEET=QUBITS_COLOUR=gold;Q(7,21)_87;####!_QUBIT_Q=88_SHEET=QUBITS_COLOUR=gold;Q(8,0)_88;####!_QUBIT_Q=89_SHEET=QUBITS_COLOUR=gold;Q(8,2)_89;####!_QUBIT_Q=90_SHEET=QUBITS_COLOUR=gold;Q(8,4)_90;####!_QUBIT_Q=91_SHEET=QUBITS_COLOUR=gold;Q(8,6)_91;####!_QUBIT_Q=92_SHEET=QUBITS_COLOUR=gold;Q(8,8)_92;####!_QUBIT_Q=93_SHEET=QUBITS_COLOUR=gold;Q(8,10)_93;####!_QUBIT_Q=94_SHEET=QUBITS_COLOUR=gold;Q(8,12)_94;####!_QUBIT_Q=95_SHEET=QUBITS_COLOUR=gold;Q(8,14)_95;####!_QUBIT_Q=96_SHEET=QUBITS_COLOUR=gold;Q(8,16)_96;####!_QUBIT_Q=97_SHEET=QUBITS_COLOUR=gold;Q(8,18)_97;####!_QUBIT_Q=98_SHEET=QUBITS_COLOUR=gold;Q(8,20)_98;####!_QUBIT_Q=99_SHEET=QUBITS_COLOUR=gold;Q(9,1)_99;####!_QUBIT_Q=100_SHEET=QUBITS_COLOUR=gold;Q(9,3)_100;####!_QUBIT_Q=101_SHEET=QUBITS_COLOUR=gold;Q(9,5)_101;####!_QUBIT_Q=102_SHEET=QUBITS_COLOUR=gold;Q(9,7)_102;####!_QUBIT_Q=103_SHEET=QUBITS_COLOUR=gold;Q(9,9)_103;####!_QUBIT_Q=104_SHEET=QUBITS_COLOUR=gold;Q(9,11)_104;####!_QUBIT_Q=105_SHEET=QUBITS_COLOUR=gold;Q(9,13)_105;####!_QUBIT_Q=106_SHEET=QUBITS_COLOUR=gold;Q(9,15)_106;####!_QUBIT_Q=107_SHEET=QUBITS_COLOUR=gold;Q(9,17)_107;####!_QUBIT_Q=108_SHEET=QUBITS_COLOUR=gold;Q(9,19)_108;####!_QUBIT_Q=109_SHEET=QUBITS_COLOUR=gold;Q(9,21)_109;####!_QUBIT_Q=110_SHEET=QUBITS_COLOUR=gold;Q(10,0)_110;####!_QUBIT_Q=111_SHEET=QUBITS_COLOUR=gold;Q(10,2)_111;####!_QUBIT_Q=112_SHEET=QUBITS_COLOUR=gold;Q(10,4)_112;####!_QUBIT_Q=113_SHEET=QUBITS_COLOUR=gold;Q(10,6)_113;####!_QUBIT_Q=114_SHEET=QUBITS_COLOUR=gold;Q(10,8)_114;####!_QUBIT_Q=115_SHEET=QUBITS_COLOUR=gold;Q(10,10)_115;####!_QUBIT_Q=116_SHEET=QUBITS_COLOUR=gold;Q(10,12)_116;####!_QUBIT_Q=117_SHEET=QUBITS_COLOUR=gold;Q(10,14)_117;####!_QUBIT_Q=118_SHEET=QUBITS_COLOUR=gold;Q(10,16)_118;####!_QUBIT_Q=119_SHEET=QUBITS_COLOUR=gold;Q(10,18)_119;####!_QUBIT_Q=120_SHEET=QUBITS_COLOUR=gold;Q(10,20)_120;####!_QUBIT_Q=121_SHEET=QUBITS_COLOUR=gold;Q(11,1)_121;####!_QUBIT_Q=122_SHEET=QUBITS_COLOUR=gold;Q(11,3)_122;####!_QUBIT_Q=123_SHEET=QUBITS_COLOUR=gold;Q(11,5)_123;####!_QUBIT_Q=124_SHEET=QUBITS_COLOUR=gold;Q(11,7)_124;####!_QUBIT_Q=125_SHEET=QUBITS_COLOUR=gold;Q(11,9)_125;####!_QUBIT_Q=126_SHEET=QUBITS_COLOUR=gold;Q(11,11)_126;####!_QUBIT_Q=127_SHEET=QUBITS_COLOUR=gold;Q(11,13)_127;####!_QUBIT_Q=128_SHEET=QUBITS_COLOUR=gold;Q(11,15)_128;####!_QUBIT_Q=129_SHEET=QUBITS_COLOUR=gold;Q(11,17)_129;####!_QUBIT_Q=130_SHEET=QUBITS_COLOUR=gold;Q(11,19)_130;####!_QUBIT_Q=131_SHEET=QUBITS_COLOUR=gold;Q(11,21)_131;####!_QUBIT_Q=132_SHEET=QUBITS_COLOUR=gold;Q(12,0)_132;####!_QUBIT_Q=133_SHEET=QUBITS_COLOUR=gold_DEFECTIVE=true;Q(12,2)_133;####!_QUBIT_Q=134_SHEET=QUBITS_COLOUR=gold;Q(12,4)_134;####!_QUBIT_Q=135_SHEET=QUBITS_COLOUR=gold;Q(12,6)_135;####!_QUBIT_Q=136_SHEET=QUBITS_COLOUR=gold;Q(12,8)_136;####!_QUBIT_Q=137_SHEET=QUBITS_COLOUR=gold;Q(12,10)_137;####!_QUBIT_Q=138_SHEET=QUBITS_COLOUR=gold;Q(12,12)_138;####!_QUBIT_Q=139_SHEET=QUBITS_COLOUR=gold;Q(12,14)_139;####!_QUBIT_Q=140_SHEET=QUBITS_COLOUR=gold;Q(12,16)_140;####!_QUBIT_Q=141_SHEET=QUBITS_COLOUR=gold;Q(12,18)_141;####!_QUBIT_Q=142_SHEET=QUBITS_COLOUR=gold;Q(12,20)_142;####!_QUBIT_Q=143_SHEET=QUBITS_COLOUR=gold;Q(13,1)_143;####!_QUBIT_Q=144_SHEET=QUBITS_COLOUR=gold;Q(13,3)_144;####!_QUBIT_Q=145_SHEET=QUBITS_COLOUR=gold;Q(13,5)_145;####!_QUBIT_Q=146_SHEET=QUBITS_COLOUR=gold;Q(13,7)_146;####!_QUBIT_Q=147_SHEET=QUBITS_COLOUR=gold;Q(13,9)_147;####!_QUBIT_Q=148_SHEET=QUBITS_COLOUR=gold;Q(13,11)_148;####!_QUBIT_Q=149_SHEET=QUBITS_COLOUR=gold;Q(13,13)_149;####!_QUBIT_Q=150_SHEET=QUBITS_COLOUR=gold;Q(13,15)_150;####!_QUBIT_Q=151_SHEET=QUBITS_COLOUR=gold;Q(13,17)_151;####!_QUBIT_Q=152_SHEET=QUBITS_COLOUR=gold;Q(13,19)_152;####!_QUBIT_Q=153_SHEET=QUBITS_COLOUR=gold;Q(13,21)_153;####!_QUBIT_Q=154_SHEET=QUBITS_COLOUR=gold;Q(14,0)_154;####!_QUBIT_Q=155_SHEET=QUBITS_COLOUR=gold;Q(14,2)_155;####!_QUBIT_Q=156_SHEET=QUBITS_COLOUR=gold;Q(14,4)_156;####!_QUBIT_Q=157_SHEET=QUBITS_COLOUR=gold;Q(14,6)_157;####!_QUBIT_Q=158_SHEET=QUBITS_COLOUR=gold;Q(14,8)_158;####!_QUBIT_Q=159_SHEET=QUBITS_COLOUR=gold;Q(14,10)_159;####!_QUBIT_Q=160_SHEET=QUBITS_COLOUR=gold;Q(14,12)_160;####!_QUBIT_Q=161_SHEET=QUBITS_COLOUR=gold;Q(14,14)_161;####!_QUBIT_Q=162_SHEET=QUBITS_COLOUR=gold;Q(14,16)_162;####!_QUBIT_Q=163_SHEET=QUBITS_COLOUR=gold;Q(14,18)_163;####!_QUBIT_Q=164_SHEET=QUBITS_COLOUR=gold;Q(14,20)_164;####!_QUBIT_Q=165_SHEET=QUBITS_COLOUR=gold;Q(15,1)_165;####!_QUBIT_Q=166_SHEET=QUBITS_COLOUR=gold;Q(15,3)_166;####!_QUBIT_Q=167_SHEET=QUBITS_COLOUR=gold;Q(15,5)_167;####!_QUBIT_Q=168_SHEET=QUBITS_COLOUR=gold;Q(15,7)_168;####!_QUBIT_Q=169_SHEET=QUBITS_COLOUR=gold;Q(15,9)_169;####!_QUBIT_Q=170_SHEET=QUBITS_COLOUR=gold;Q(15,11)_170;####!_QUBIT_Q=171_SHEET=QUBITS_COLOUR=gold;Q(15,13)_171;####!_QUBIT_Q=172_SHEET=QUBITS_COLOUR=gold;Q(15,15)_172;####!_QUBIT_Q=173_SHEET=QUBITS_COLOUR=gold;Q(15,17)_173;####!_QUBIT_Q=174_SHEET=QUBITS_COLOUR=gold;Q(15,19)_174;####!_QUBIT_Q=175_SHEET=QUBITS_COLOUR=gold;Q(15,21)_175;####!_QUBIT_Q=176_SHEET=QUBITS_COLOUR=gold;Q(16,0)_176;####!_QUBIT_Q=177_SHEET=QUBITS_COLOUR=gold;Q(16,2)_177;####!_QUBIT_Q=178_SHEET=QUBITS_COLOUR=gold;Q(16,4)_178;####!_QUBIT_Q=179_SHEET=QUBITS_COLOUR=gold;Q(16,6)_179;####!_QUBIT_Q=180_SHEET=QUBITS_COLOUR=gold;Q(16,8)_180;####!_QUBIT_Q=181_SHEET=QUBITS_COLOUR=gold;Q(16,10)_181;####!_QUBIT_Q=182_SHEET=QUBITS_COLOUR=gold;Q(16,12)_182;####!_QUBIT_Q=183_SHEET=QUBITS_COLOUR=gold;Q(16,14)_183;####!_QUBIT_Q=184_SHEET=QUBITS_COLOUR=gold;Q(16,16)_184;####!_QUBIT_Q=185_SHEET=QUBITS_COLOUR=gold;Q(16,18)_185;####!_QUBIT_Q=186_SHEET=QUBITS_COLOUR=gold;Q(16,20)_186;####!_QUBIT_Q=187_SHEET=QUBITS_COLOUR=gold;Q(17,1)_187;####!_QUBIT_Q=188_SHEET=QUBITS_COLOUR=gold;Q(17,3)_188;####!_QUBIT_Q=189_SHEET=QUBITS_COLOUR=gold;Q(17,5)_189;####!_QUBIT_Q=190_SHEET=QUBITS_COLOUR=gold;Q(17,7)_190;####!_QUBIT_Q=191_SHEET=QUBITS_COLOUR=gold;Q(17,9)_191;####!_QUBIT_Q=192_SHEET=QUBITS_COLOUR=gold;Q(17,11)_192;####!_QUBIT_Q=193_SHEET=QUBITS_COLOUR=gold;Q(17,13)_193;####!_QUBIT_Q=194_SHEET=QUBITS_COLOUR=gold;Q(17,15)_194;####!_QUBIT_Q=195_SHEET=QUBITS_COLOUR=gold;Q(17,17)_195;####!_QUBIT_Q=196_SHEET=QUBITS_COLOUR=gold;Q(17,19)_196;####!_QUBIT_Q=197_SHEET=QUBITS_COLOUR=gold;Q(17,21)_197;####!_QUBIT_Q=198_SHEET=QUBITS_COLOUR=gold;Q(18,0)_198;####!_QUBIT_Q=199_SHEET=QUBITS_COLOUR=gold;Q(18,2)_199;####!_QUBIT_Q=200_SHEET=QUBITS_COLOUR=gold;Q(18,4)_200;####!_QUBIT_Q=201_SHEET=QUBITS_COLOUR=gold;Q(18,6)_201;####!_QUBIT_Q=202_SHEET=QUBITS_COLOUR=gold;Q(18,8)_202;####!_QUBIT_Q=203_SHEET=QUBITS_COLOUR=gold;Q(18,10)_203;####!_QUBIT_Q=204_SHEET=QUBITS_COLOUR=gold;Q(18,12)_204;####!_QUBIT_Q=205_SHEET=QUBITS_COLOUR=gold;Q(18,14)_205;####!_QUBIT_Q=206_SHEET=QUBITS_COLOUR=gold;Q(18,16)_206;####!_QUBIT_Q=207_SHEET=QUBITS_COLOUR=gold;Q(18,18)_207;####!_QUBIT_Q=208_SHEET=QUBITS_COLOUR=gold;Q(18,20)_208;####!_QUBIT_Q=209_SHEET=QUBITS_COLOUR=gold;Q(19,1)_209;####!_QUBIT_Q=210_SHEET=QUBITS_COLOUR=gold;Q(19,3)_210;####!_QUBIT_Q=211_SHEET=QUBITS_COLOUR=gold;Q(19,5)_211;####!_QUBIT_Q=212_SHEET=QUBITS_COLOUR=gold;Q(19,7)_212;####!_QUBIT_Q=213_SHEET=QUBITS_COLOUR=gold;Q(19,9)_213;####!_QUBIT_Q=214_SHEET=QUBITS_COLOUR=gold;Q(19,11)_214;####!_QUBIT_Q=215_SHEET=QUBITS_COLOUR=gold;Q(19,13)_215;####!_QUBIT_Q=216_SHEET=QUBITS_COLOUR=gold;Q(19,15)_216;####!_QUBIT_Q=217_SHEET=QUBITS_COLOUR=gold;Q(19,17)_217;####!_QUBIT_Q=218_SHEET=QUBITS_COLOUR=gold;Q(19,19)_218;####!_QUBIT_Q=219_SHEET=QUBITS_COLOUR=gold;Q(19,21)_219;####!_QUBIT_Q=220_SHEET=QUBITS_COLOUR=gold;Q(20,0)_220;####!_QUBIT_Q=221_SHEET=QUBITS_COLOUR=gold;Q(20,2)_221;####!_QUBIT_Q=222_SHEET=QUBITS_COLOUR=gold;Q(20,4)_222;####!_QUBIT_Q=223_SHEET=QUBITS_COLOUR=gold;Q(20,6)_223;####!_QUBIT_Q=224_SHEET=QUBITS_COLOUR=gold;Q(20,8)_224;####!_QUBIT_Q=225_SHEET=QUBITS_COLOUR=gold;Q(20,10)_225;####!_QUBIT_Q=226_SHEET=QUBITS_COLOUR=gold;Q(20,12)_226;####!_QUBIT_Q=227_SHEET=QUBITS_COLOUR=gold;Q(20,14)_227;####!_QUBIT_Q=228_SHEET=QUBITS_COLOUR=gold;Q(20,16)_228;####!_QUBIT_Q=229_SHEET=QUBITS_COLOUR=gold;Q(20,18)_229;####!_QUBIT_Q=230_SHEET=QUBITS_COLOUR=gold;Q(20,20)_230;####!_QUBIT_Q=231_SHEET=QUBITS_COLOUR=gold;Q(21,1)_231;####!_QUBIT_Q=232_SHEET=QUBITS_COLOUR=gold;Q(21,3)_232;####!_QUBIT_Q=233_SHEET=QUBITS_COLOUR=gold;Q(21,5)_233;####!_QUBIT_Q=234_SHEET=QUBITS_COLOUR=gold;Q(21,7)_234;####!_QUBIT_Q=235_SHEET=QUBITS_COLOUR=gold;Q(21,9)_235;####!_QUBIT_Q=236_SHEET=QUBITS_COLOUR=gold;Q(21,11)_236;####!_QUBIT_Q=237_SHEET=QUBITS_COLOUR=gold;Q(21,13)_237;####!_QUBIT_Q=238_SHEET=QUBITS_COLOUR=gold;Q(21,15)_238;####!_QUBIT_Q=239_SHEET=QUBITS_COLOUR=gold;Q(21,17)_239;####!_QUBIT_Q=240_SHEET=QUBITS_COLOUR=gold;Q(21,19)_240;Q(22,0)_241;Q(22,1)_242;####!_HIGHLIGHT_TARGET=QUBIT_QUBITS=52_COLOR=red;####!_HIGHLIGHT_TARGET=QUBIT_QUBITS=133_COLOR=red;####!_CONN_SET_EDGES=(0-11,1-12,1-11,2-13,2-12,3-14,3-13,4-15,4-14,5-16,5-15,6-17,6-16,7-18,7-17,8-19,8-18,9-20,9-19,10-21,10-20,11-22,12-23,13-24,14-25,15-26,16-27,17-28,18-29,19-30,20-31,21-32,22-33,23-34,23-33,24-35,24-34,25-36,25-35,26-37,26-36,27-38,27-37,28-39,28-38,29-40,29-39,30-41,30-40,31-42,31-41,32-43,32-42,33-44,34-45,35-46,36-47,37-48,38-49,39-50,40-51,41-52,42-53,43-54,44-55,45-56,45-55,46-57,46-56,47-58,47-57,48-59,48-58,49-60,49-59,50-61,50-60,51-62,51-61,52-63,52-62,53-64,53-63,54-65,54-64,55-66,56-67,57-68,58-69,59-70,60-71,61-72,62-73,63-74,64-75,65-76,66-77,67-78,67-77,68-79,68-78,69-80,69-79,70-81,70-80,71-82,71-81,72-83,72-82,73-84,73-83,74-85,74-84,75-85,76-87,76-86,77-88,78-89,79-90,80-91,81-92,82-93,83-94,84-95,85-96,86-97,87-98,88-99,89-100,89-99,90-101,90-100,91-102,91-101,92-103,92-102,93-104,93-103,94-105,94-104,95-106,95-105,96-107,96-106,97-108,97-107,98-109,98-108,99-110,100-111,101-112,102-113,103-114,104-115,105-116,106-117,107-118,108-119,109-120,110-121,111-122,111-121,112-123,112-122,113-124,113-123,114-125,114-124,115-126,115-125,116-127,116-126,117-128,117-127,118-129,118-128,119-130,119-129,120-131,120-130,121-132,122-133,123-134,124-135,125-136,126-137,127-138,128-139,129-140,130-141,131-142,132-143,133-144,133-143,134-145,134-144,135-146,135-145,136-147,136-146,137-148,137-147,138-149,138-148,139-150,139-149,140-151,140-150,141-152,141-151,142-153,142-152,143-154,144-155,145-156,146-157,147-158,148-159,149-160,150-161,151-162,152-163,153-164,154-165,155-166,155-165,156-167,156-166,157-168,157-167,158-169,158-168,159-170,159-169,160-171,160-170,161-172,161-171,162-173,162-172,163-174,163-173,164-175,164-174,165-176,166-177,167-178,168-179,169-180,170-181,171-182,172-183,173-184,174-185,175-186,176-187,177-188,177-187,178-189,178-188,179-190,179-189,180-191,180-190,181-192,181-191,182-193,182-192,183-194,183-193,184-195,184-194,185-196,185-195,186-197,186-196,187-198,188-199,189-200,190-201,191-202,192-203,193-204,194-205,195-206,196-207,197-208,198-209,199-210,199-209,200-211,200-210,201-212,201-211,202-213,202-212,203-214,203-213,204-215,204-214,205-216,205-215,206-217,206-216,207-218,207-217,208-219,208-218,209-220,210-221,211-222,212-223,213-224,214-225,215-226,216-227,217-228,218-229,219-230,220-231,221-232,221-231,222-233,222-232,223-234,223-233,224-235,224-234,225-236,225-235,226-237,226-236,227-238,227-237,228-239,228-238,229-240,229-239,230-240)_SHEET=E_COLOUR=##4361ee;####!_CONN_SET_EDGES=(75-86)_SHEET=E_COLOUR=red;####!_POLY_SHEET=UNTX_STROKE=none_FILL=(1,0,0,0.15);##!pragma_POLYGON(1,0,0,0.15)_1_11_0;####!_POLY_SHEET=UNTX_STROKE=none_FILL=(1,0,0,0.15);##!pragma_POLYGON(1,0,0,0.15)_2_12_1;####!_POLY_SHEET=UNTX_STROKE=none_FILL=(1,0,0,0.15);##!pragma_POLYGON(1,0,0,0.15)_3_13_2;####!_POLY_SHEET=UNTX_STROKE=none_FILL=(1,0,0,0.15);##!pragma_POLYGON(1,0,0,0.15)_4_14_3;####!_POLY_SHEET=UNTX_STROKE=none_FILL=(1,0,0,0.15);##!pragma_POLYGON(1,0,0,0.15)_5_15_4;####!_POLY_SHEET=UNTX_STROKE=none_FILL=(1,0,0,0.15);##!pragma_POLYGON(1,0,0,0.15)_6_16_5;####!_POLY_SHEET=UNTX_STROKE=none_FILL=(1,0,0,0.15);##!pragma_POLYGON(1,0,0,0.15)_7_17_6;####!_POLY_SHEET=UNTX_STROKE=none_FILL=(1,0,0,0.15);##!pragma_POLYGON(1,0,0,0.15)_8_18_7;####!_POLY_SHEET=UNTX_STROKE=none_FILL=(1,0,0,0.15);##!pragma_POLYGON(1,0,0,0.15)_9_19_8;####!_POLY_SHEET=UNTX_STROKE=none_FILL=(1,0,0,0.15);##!pragma_POLYGON(1,0,0,0.15)_10_20_9;####!_POLY_SHEET=UNTX_STROKE=none_FILL=(1,0,0,0.15);##!pragma_POLYGON(1,0,0,0.15)_11_23_33_22;####!_POLY_SHEET=UNTX_STROKE=none_FILL=(1,0,0,0.15);##!pragma_POLYGON(1,0,0,0.15)_12_24_34_23;####!_POLY_SHEET=UNTX_STROKE=none_FILL=(1,0,0,0.15);##!pragma_POLYGON(1,0,0,0.15)_13_25_35_24;####!_POLY_SHEET=UNTX_STROKE=none_FILL=(1,0,0,0.15);##!pragma_POLYGON(1,0,0,0.15)_14_26_36_25;####!_POLY_SHEET=UNTX_STROKE=none_FILL=(1,0,0,0.15);##!pragma_POLYGON(1,0,0,0.15)_15_27_37_26;####!_POLY_SHEET=UNTX_STROKE=none_FILL=(1,0,0,0.15);##!pragma_POLYGON(1,0,0,0.15)_16_28_38_27;####!_POLY_SHEET=UNTX_STROKE=none_FILL=(1,0,0,0.15);##!pragma_POLYGON(1,0,0,0.15)_17_29_39_28;####!_POLY_SHEET=UNTX_STROKE=none_FILL=(1,0,0,0.15);##!pragma_POLYGON(1,0,0,0.15)_18_30_40_29;####!_POLY_SHEET=UNTX_STROKE=none_FILL=(1,0,0,0.15);##!pragma_POLYGON(1,0,0,0.15)_19_31_41_30;####!_POLY_SHEET=UNTX_STROKE=none_FILL=(1,0,0,0.15);##!pragma_POLYGON(1,0,0,0.15)_20_32_42_31;####!_POLY_SHEET=UNTX_STROKE=none_FILL=(1,0,0,0.15);##!pragma_POLYGON(1,0,0,0.15)_33_45_55_44;####!_POLY_SHEET=UNTX_STROKE=none_FILL=(1,0,0,0.15);##!pragma_POLYGON(1,0,0,0.15)_34_46_56_45;####!_POLY_SHEET=UNTX_STROKE=none_FILL=(1,0,0,0.15);##!pragma_POLYGON(1,0,0,0.15)_35_47_57_46;####!_POLY_SHEET=UNTX_STROKE=none_FILL=(1,0,0,0.15);##!pragma_POLYGON(1,0,0,0.15)_36_48_58_47;####!_POLY_SHEET=UNTX_STROKE=none_FILL=(1,0,0,0.15);##!pragma_POLYGON(1,0,0,0.15)_37_49_59_48;####!_POLY_SHEET=UNTX_STROKE=none_FILL=(1,0,0,0.15);##!pragma_POLYGON(1,0,0,0.15)_38_50_60_49;####!_POLY_SHEET=UNTX_STROKE=none_FILL=(1,0,0,0.15);##!pragma_POLYGON(1,0,0,0.15)_39_51_61_50;####!_POLY_SHEET=UNTX_STROKE=none_FILL=(1,0,0,0.15);##!pragma_POLYGON(1,0,0,0.15)_42_54_64_53;####!_POLY_SHEET=UNTX_STROKE=none_FILL=(1,0,0,0.15);##!pragma_POLYGON(1,0,0,0.15)_55_67_77_66;####!_POLY_SHEET=UNTX_STROKE=none_FILL=(1,0,0,0.15);##!pragma_POLYGON(1,0,0,0.15)_56_68_78_67;####!_POLY_SHEET=UNTX_STROKE=none_FILL=(1,0,0,0.15);##!pragma_POLYGON(1,0,0,0.15)_57_69_79_68;####!_POLY_SHEET=UNTX_STROKE=none_FILL=(1,0,0,0.15);##!pragma_POLYGON(1,0,0,0.15)_58_70_80_69;####!_POLY_SHEET=UNTX_STROKE=none_FILL=(1,0,0,0.15);##!pragma_POLYGON(1,0,0,0.15)_59_71_81_70;####!_POLY_SHEET=UNTX_STROKE=none_FILL=(1,0,0,0.15);##!pragma_POLYGON(1,0,0,0.15)_60_72_82_71;####!_POLY_SHEET=UNTX_STROKE=none_FILL=(1,0,0,0.15);##!pragma_POLYGON(1,0,0,0.15)_61_73_83_72;####!_POLY_SHEET=UNTX_STROKE=none_FILL=(1,0,0,0.15);##!pragma_POLYGON(1,0,0,0.15)_63_75_85_74;####!_POLY_SHEET=UNTX_STROKE=none_FILL=(1,0,0,0.15);##!pragma_POLYGON(1,0,0,0.15)_77_89_99_88;####!_POLY_SHEET=UNTX_STROKE=none_FILL=(1,0,0,0.15);##!pragma_POLYGON(1,0,0,0.15)_78_90_100_89;####!_POLY_SHEET=UNTX_STROKE=none_FILL=(1,0,0,0.15);##!pragma_POLYGON(1,0,0,0.15)_79_91_101_90;####!_POLY_SHEET=UNTX_STROKE=none_FILL=(1,0,0,0.15);##!pragma_POLYGON(1,0,0,0.15)_80_92_102_91;####!_POLY_SHEET=UNTX_STROKE=none_FILL=(1,0,0,0.15);##!pragma_POLYGON(1,0,0,0.15)_81_93_103_92;####!_POLY_SHEET=UNTX_STROKE=none_FILL=(1,0,0,0.15);##!pragma_POLYGON(1,0,0,0.15)_82_94_104_93;####!_POLY_SHEET=UNTX_STROKE=none_FILL=(1,0,0,0.15);##!pragma_POLYGON(1,0,0,0.15)_83_95_105_94;####!_POLY_SHEET=UNTX_STROKE=none_FILL=(1,0,0,0.15);##!pragma_POLYGON(1,0,0,0.15)_84_96_106_95;####!_POLY_SHEET=UNTX_STROKE=none_FILL=(1,0,0,0.15);##!pragma_POLYGON(1,0,0,0.15)_86_98_108_97;####!_POLY_SHEET=UNTX_STROKE=none_FILL=(1,0,0,0.15);##!pragma_POLYGON(1,0,0,0.15)_99_111_121_110;####!_POLY_SHEET=UNTX_STROKE=none_FILL=(1,0,0,0.15);##!pragma_POLYGON(1,0,0,0.15)_100_112_122_111;####!_POLY_SHEET=UNTX_STROKE=none_FILL=(1,0,0,0.15);##!pragma_POLYGON(1,0,0,0.15)_101_113_123_112;####!_POLY_SHEET=UNTX_STROKE=none_FILL=(1,0,0,0.15);##!pragma_POLYGON(1,0,0,0.15)_102_114_124_113;####!_POLY_SHEET=UNTX_STROKE=none_FILL=(1,0,0,0.15);##!pragma_POLYGON(1,0,0,0.15)_103_115_125_114;####!_POLY_SHEET=UNTX_STROKE=none_FILL=(1,0,0,0.15);##!pragma_POLYGON(1,0,0,0.15)_104_116_126_115;####!_POLY_SHEET=UNTX_STROKE=none_FILL=(1,0,0,0.15);##!pragma_POLYGON(1,0,0,0.15)_105_117_127_116;####!_POLY_SHEET=UNTX_STROKE=none_FILL=(1,0,0,0.15);##!pragma_POLYGON(1,0,0,0.15)_106_118_128_117;####!_POLY_SHEET=UNTX_STROKE=none_FILL=(1,0,0,0.15);##!pragma_POLYGON(1,0,0,0.15)_107_119_129_118;####!_POLY_SHEET=UNTX_STROKE=none_FILL=(1,0,0,0.15);##!pragma_POLYGON(1,0,0,0.15)_108_120_130_119;####!_POLY_SHEET=UNTX_STROKE=none_FILL=(1,0,0,0.15);##!pragma_POLYGON(1,0,0,0.15)_123_135_145_134;####!_POLY_SHEET=UNTX_STROKE=none_FILL=(1,0,0,0.15);##!pragma_POLYGON(1,0,0,0.15)_124_136_146_135;####!_POLY_SHEET=UNTX_STROKE=none_FILL=(1,0,0,0.15);##!pragma_POLYGON(1,0,0,0.15)_125_137_147_136;####!_POLY_SHEET=UNTX_STROKE=none_FILL=(1,0,0,0.15);##!pragma_POLYGON(1,0,0,0.15)_126_138_148_137;####!_POLY_SHEET=UNTX_STROKE=none_FILL=(1,0,0,0.15);##!pragma_POLYGON(1,0,0,0.15)_127_139_149_138;####!_POLY_SHEET=UNTX_STROKE=none_FILL=(1,0,0,0.15);##!pragma_POLYGON(1,0,0,0.15)_128_140_150_139;####!_POLY_SHEET=UNTX_STROKE=none_FILL=(1,0,0,0.15);##!pragma_POLYGON(1,0,0,0.15)_129_141_151_140;####!_POLY_SHEET=UNTX_STROKE=none_FILL=(1,0,0,0.15);##!pragma_POLYGON(1,0,0,0.15)_130_142_152_141;####!_POLY_SHEET=UNTX_STROKE=none_FILL=(1,0,0,0.15);##!pragma_POLYGON(1,0,0,0.15)_144_156_166_155;####!_POLY_SHEET=UNTX_STROKE=none_FILL=(1,0,0,0.15);##!pragma_POLYGON(1,0,0,0.15)_145_157_167_156;####!_POLY_SHEET=UNTX_STROKE=none_FILL=(1,0,0,0.15);##!pragma_POLYGON(1,0,0,0.15)_146_158_168_157;####!_POLY_SHEET=UNTX_STROKE=none_FILL=(1,0,0,0.15);##!pragma_POLYGON(1,0,0,0.15)_147_159_169_158;####!_POLY_SHEET=UNTX_STROKE=none_FILL=(1,0,0,0.15);##!pragma_POLYGON(1,0,0,0.15)_148_160_170_159;####!_POLY_SHEET=UNTX_STROKE=none_FILL=(1,0,0,0.15);##!pragma_POLYGON(1,0,0,0.15)_149_161_171_160;####!_POLY_SHEET=UNTX_STROKE=none_FILL=(1,0,0,0.15);##!pragma_POLYGON(1,0,0,0.15)_150_162_172_161;####!_POLY_SHEET=UNTX_STROKE=none_FILL=(1,0,0,0.15);##!pragma_POLYGON(1,0,0,0.15)_151_163_173_162;####!_POLY_SHEET=UNTX_STROKE=none_FILL=(1,0,0,0.15);##!pragma_POLYGON(1,0,0,0.15)_152_164_174_163;####!_POLY_SHEET=UNTX_STROKE=none_FILL=(1,0,0,0.15);##!pragma_POLYGON(1,0,0,0.15)_165_177_187_176;####!_POLY_SHEET=UNTX_STROKE=none_FILL=(1,0,0,0.15);##!pragma_POLYGON(1,0,0,0.15)_166_178_188_177;####!_POLY_SHEET=UNTX_STROKE=none_FILL=(1,0,0,0.15);##!pragma_POLYGON(1,0,0,0.15)_167_179_189_178;####!_POLY_SHEET=UNTX_STROKE=none_FILL=(1,0,0,0.15);##!pragma_POLYGON(1,0,0,0.15)_168_180_190_179;####!_POLY_SHEET=UNTX_STROKE=none_FILL=(1,0,0,0.15);##!pragma_POLYGON(1,0,0,0.15)_169_181_191_180;####!_POLY_SHEET=UNTX_STROKE=none_FILL=(1,0,0,0.15);##!pragma_POLYGON(1,0,0,0.15)_170_182_192_181;####!_POLY_SHEET=UNTX_STROKE=none_FILL=(1,0,0,0.15);##!pragma_POLYGON(1,0,0,0.15)_171_183_193_182;####!_POLY_SHEET=UNTX_STROKE=none_FILL=(1,0,0,0.15);##!pragma_POLYGON(1,0,0,0.15)_172_184_194_183;####!_POLY_SHEET=UNTX_STROKE=none_FILL=(1,0,0,0.15);##!pragma_POLYGON(1,0,0,0.15)_173_185_195_184;####!_POLY_SHEET=UNTX_STROKE=none_FILL=(1,0,0,0.15);##!pragma_POLYGON(1,0,0,0.15)_174_186_196_185;####!_POLY_SHEET=UNTX_STROKE=none_FILL=(1,0,0,0.15);##!pragma_POLYGON(1,0,0,0.15)_187_199_209_198;####!_POLY_SHEET=UNTX_STROKE=none_FILL=(1,0,0,0.15);##!pragma_POLYGON(1,0,0,0.15)_188_200_210_199;####!_POLY_SHEET=UNTX_STROKE=none_FILL=(1,0,0,0.15);##!pragma_POLYGON(1,0,0,0.15)_189_201_211_200;####!_POLY_SHEET=UNTX_STROKE=none_FILL=(1,0,0,0.15);##!pragma_POLYGON(1,0,0,0.15)_190_202_212_201;####!_POLY_SHEET=UNTX_STROKE=none_FILL=(1,0,0,0.15);##!pragma_POLYGON(1,0,0,0.15)_191_203_213_202;####!_POLY_SHEET=UNTX_STROKE=none_FILL=(1,0,0,0.15);##!pragma_POLYGON(1,0,0,0.15)_192_204_214_203;####!_POLY_SHEET=UNTX_STROKE=none_FILL=(1,0,0,0.15);##!pragma_POLYGON(1,0,0,0.15)_193_205_215_204;####!_POLY_SHEET=UNTX_STROKE=none_FILL=(1,0,0,0.15);##!pragma_POLYGON(1,0,0,0.15)_194_206_216_205;####!_POLY_SHEET=UNTX_STROKE=none_FILL=(1,0,0,0.15);##!pragma_POLYGON(1,0,0,0.15)_195_207_217_206;####!_POLY_SHEET=UNTX_STROKE=none_FILL=(1,0,0,0.15);##!pragma_POLYGON(1,0,0,0.15)_196_208_218_207;####!_POLY_SHEET=UNTX_STROKE=none_FILL=(1,0,0,0.15);##!pragma_POLYGON(1,0,0,0.15)_209_221_231_220;####!_POLY_SHEET=UNTX_STROKE=none_FILL=(1,0,0,0.15);##!pragma_POLYGON(1,0,0,0.15)_210_222_232_221;####!_POLY_SHEET=UNTX_STROKE=none_FILL=(1,0,0,0.15);##!pragma_POLYGON(1,0,0,0.15)_211_223_233_222;####!_POLY_SHEET=UNTX_STROKE=none_FILL=(1,0,0,0.15);##!pragma_POLYGON(1,0,0,0.15)_212_224_234_223;####!_POLY_SHEET=UNTX_STROKE=none_FILL=(1,0,0,0.15);##!pragma_POLYGON(1,0,0,0.15)_213_225_235_224;####!_POLY_SHEET=UNTX_STROKE=none_FILL=(1,0,0,0.15);##!pragma_POLYGON(1,0,0,0.15)_214_226_236_225;####!_POLY_SHEET=UNTX_STROKE=none_FILL=(1,0,0,0.15);##!pragma_POLYGON(1,0,0,0.15)_215_227_237_226;####!_POLY_SHEET=UNTX_STROKE=none_FILL=(1,0,0,0.15);##!pragma_POLYGON(1,0,0,0.15)_216_228_238_227;####!_POLY_SHEET=UNTX_STROKE=none_FILL=(1,0,0,0.15);##!pragma_POLYGON(1,0,0,0.15)_217_229_239_228;####!_POLY_SHEET=UNTX_STROKE=none_FILL=(1,0,0,0.15);##!pragma_POLYGON(1,0,0,0.15)_218_230_240_229;####!_POLY_SHEET=UNTX_STROKE=none_FILL=(1,0,0,0.15);##!pragma_POLYGON(1,0,0,0.15)_231;####!_POLY_SHEET=UNTX_STROKE=none_FILL=(1,0,0,0.15);##!pragma_POLYGON(1,0,0,0.15)_232;####!_POLY_SHEET=UNTX_STROKE=none_FILL=(1,0,0,0.15);##!pragma_POLYGON(1,0,0,0.15)_233;####!_POLY_SHEET=UNTX_STROKE=none_FILL=(1,0,0,0.15);##!pragma_POLYGON(1,0,0,0.15)_234;####!_POLY_SHEET=UNTX_STROKE=none_FILL=(1,0,0,0.15);##!pragma_POLYGON(1,0,0,0.15)_235;####!_POLY_SHEET=UNTX_STROKE=none_FILL=(1,0,0,0.15);##!pragma_POLYGON(1,0,0,0.15)_236;####!_POLY_SHEET=UNTX_STROKE=none_FILL=(1,0,0,0.15);##!pragma_POLYGON(1,0,0,0.15)_237;####!_POLY_SHEET=UNTX_STROKE=none_FILL=(1,0,0,0.15);##!pragma_POLYGON(1,0,0,0.15)_238;####!_POLY_SHEET=UNTX_STROKE=none_FILL=(1,0,0,0.15);##!pragma_POLYGON(1,0,0,0.15)_239;####!_POLY_SHEET=UNTX_STROKE=none_FILL=(1,0,0,0.15);##!pragma_POLYGON(1,0,0,0.15)_240;####!_POLY_SHEET=UNTZ_STROKE=none_FILL=(0,0,1,0.15);##!pragma_POLYGON(0,0,1,0.15)_11_22_0;####!_POLY_SHEET=UNTZ_STROKE=none_FILL=(0,0,1,0.15);##!pragma_POLYGON(0,0,1,0.15)_1_12_23_11;####!_POLY_SHEET=UNTZ_STROKE=none_FILL=(0,0,1,0.15);##!pragma_POLYGON(0,0,1,0.15)_2_13_24_12;####!_POLY_SHEET=UNTZ_STROKE=none_FILL=(0,0,1,0.15);##!pragma_POLYGON(0,0,1,0.15)_3_14_25_13;####!_POLY_SHEET=UNTZ_STROKE=none_FILL=(0,0,1,0.15);##!pragma_POLYGON(0,0,1,0.15)_4_15_26_14;####!_POLY_SHEET=UNTZ_STROKE=none_FILL=(0,0,1,0.15);##!pragma_POLYGON(0,0,1,0.15)_5_16_27_15;####!_POLY_SHEET=UNTZ_STROKE=none_FILL=(0,0,1,0.15);##!pragma_POLYGON(0,0,1,0.15)_6_17_28_16;####!_POLY_SHEET=UNTZ_STROKE=none_FILL=(0,0,1,0.15);##!pragma_POLYGON(0,0,1,0.15)_7_18_29_17;####!_POLY_SHEET=UNTZ_STROKE=none_FILL=(0,0,1,0.15);##!pragma_POLYGON(0,0,1,0.15)_8_19_30_18;####!_POLY_SHEET=UNTZ_STROKE=none_FILL=(0,0,1,0.15);##!pragma_POLYGON(0,0,1,0.15)_9_20_31_19;####!_POLY_SHEET=UNTZ_STROKE=none_FILL=(0,0,1,0.15);##!pragma_POLYGON(0,0,1,0.15)_10_21_32_20;####!_POLY_SHEET=UNTZ_STROKE=none_FILL=(0,0,1,0.15);##!pragma_POLYGON(0,0,1,0.15)_33_44_22;####!_POLY_SHEET=UNTZ_STROKE=none_FILL=(0,0,1,0.15);##!pragma_POLYGON(0,0,1,0.15)_23_34_45_33;####!_POLY_SHEET=UNTZ_STROKE=none_FILL=(0,0,1,0.15);##!pragma_POLYGON(0,0,1,0.15)_24_35_46_34;####!_POLY_SHEET=UNTZ_STROKE=none_FILL=(0,0,1,0.15);##!pragma_POLYGON(0,0,1,0.15)_25_36_47_35;####!_POLY_SHEET=UNTZ_STROKE=none_FILL=(0,0,1,0.15);##!pragma_POLYGON(0,0,1,0.15)_26_37_48_36;####!_POLY_SHEET=UNTZ_STROKE=none_FILL=(0,0,1,0.15);##!pragma_POLYGON(0,0,1,0.15)_27_38_49_37;####!_POLY_SHEET=UNTZ_STROKE=none_FILL=(0,0,1,0.15);##!pragma_POLYGON(0,0,1,0.15)_28_39_50_38;####!_POLY_SHEET=UNTZ_STROKE=none_FILL=(0,0,1,0.15);##!pragma_POLYGON(0,0,1,0.15)_29_40_51_39;####!_POLY_SHEET=UNTZ_STROKE=none_FILL=(0,0,1,0.15);##!pragma_POLYGON(0,0,1,0.15)_32_43_54_42;####!_POLY_SHEET=UNTZ_STROKE=none_FILL=(0,0,1,0.15);##!pragma_POLYGON(0,0,1,0.15)_55_66_44;####!_POLY_SHEET=UNTZ_STROKE=none_FILL=(0,0,1,0.15);##!pragma_POLYGON(0,0,1,0.15)_45_56_67_55;####!_POLY_SHEET=UNTZ_STROKE=none_FILL=(0,0,1,0.15);##!pragma_POLYGON(0,0,1,0.15)_46_57_68_56;####!_POLY_SHEET=UNTZ_STROKE=none_FILL=(0,0,1,0.15);##!pragma_POLYGON(0,0,1,0.15)_47_58_69_57;####!_POLY_SHEET=UNTZ_STROKE=none_FILL=(0,0,1,0.15);##!pragma_POLYGON(0,0,1,0.15)_48_59_70_58;####!_POLY_SHEET=UNTZ_STROKE=none_FILL=(0,0,1,0.15);##!pragma_POLYGON(0,0,1,0.15)_49_60_71_59;####!_POLY_SHEET=UNTZ_STROKE=none_FILL=(0,0,1,0.15);##!pragma_POLYGON(0,0,1,0.15)_50_61_72_60;####!_POLY_SHEET=UNTZ_STROKE=none_FILL=(0,0,1,0.15);##!pragma_POLYGON(0,0,1,0.15)_51_62_73_61;####!_POLY_SHEET=UNTZ_STROKE=none_FILL=(0,0,1,0.15);##!pragma_POLYGON(0,0,1,0.15)_53_64_75_63;####!_POLY_SHEET=UNTZ_STROKE=none_FILL=(0,0,1,0.15);##!pragma_POLYGON(0,0,1,0.15)_77_88_66;####!_POLY_SHEET=UNTZ_STROKE=none_FILL=(0,0,1,0.15);##!pragma_POLYGON(0,0,1,0.15)_67_78_89_77;####!_POLY_SHEET=UNTZ_STROKE=none_FILL=(0,0,1,0.15);##!pragma_POLYGON(0,0,1,0.15)_68_79_90_78;####!_POLY_SHEET=UNTZ_STROKE=none_FILL=(0,0,1,0.15);##!pragma_POLYGON(0,0,1,0.15)_69_80_91_79;####!_POLY_SHEET=UNTZ_STROKE=none_FILL=(0,0,1,0.15);##!pragma_POLYGON(0,0,1,0.15)_70_81_92_80;####!_POLY_SHEET=UNTZ_STROKE=none_FILL=(0,0,1,0.15);##!pragma_POLYGON(0,0,1,0.15)_71_82_93_81;####!_POLY_SHEET=UNTZ_STROKE=none_FILL=(0,0,1,0.15);##!pragma_POLYGON(0,0,1,0.15)_72_83_94_82;####!_POLY_SHEET=UNTZ_STROKE=none_FILL=(0,0,1,0.15);##!pragma_POLYGON(0,0,1,0.15)_73_84_95_83;####!_POLY_SHEET=UNTZ_STROKE=none_FILL=(0,0,1,0.15);##!pragma_POLYGON(0,0,1,0.15)_74_85_96_84;####!_POLY_SHEET=UNTZ_STROKE=none_FILL=(0,0,1,0.15);##!pragma_POLYGON(0,0,1,0.15)_76_87_98_86;####!_POLY_SHEET=UNTZ_STROKE=none_FILL=(0,0,1,0.15);##!pragma_POLYGON(0,0,1,0.15)_99_110_88;####!_POLY_SHEET=UNTZ_STROKE=none_FILL=(0,0,1,0.15);##!pragma_POLYGON(0,0,1,0.15)_89_100_111_99;####!_POLY_SHEET=UNTZ_STROKE=none_FILL=(0,0,1,0.15);##!pragma_POLYGON(0,0,1,0.15)_90_101_112_100;####!_POLY_SHEET=UNTZ_STROKE=none_FILL=(0,0,1,0.15);##!pragma_POLYGON(0,0,1,0.15)_91_102_113_101;####!_POLY_SHEET=UNTZ_STROKE=none_FILL=(0,0,1,0.15);##!pragma_POLYGON(0,0,1,0.15)_92_103_114_102;####!_POLY_SHEET=UNTZ_STROKE=none_FILL=(0,0,1,0.15);##!pragma_POLYGON(0,0,1,0.15)_93_104_115_103;####!_POLY_SHEET=UNTZ_STROKE=none_FILL=(0,0,1,0.15);##!pragma_POLYGON(0,0,1,0.15)_94_105_116_104;####!_POLY_SHEET=UNTZ_STROKE=none_FILL=(0,0,1,0.15);##!pragma_POLYGON(0,0,1,0.15)_95_106_117_105;####!_POLY_SHEET=UNTZ_STROKE=none_FILL=(0,0,1,0.15);##!pragma_POLYGON(0,0,1,0.15)_96_107_118_106;####!_POLY_SHEET=UNTZ_STROKE=none_FILL=(0,0,1,0.15);##!pragma_POLYGON(0,0,1,0.15)_97_108_119_107;####!_POLY_SHEET=UNTZ_STROKE=none_FILL=(0,0,1,0.15);##!pragma_POLYGON(0,0,1,0.15)_98_109_120_108;####!_POLY_SHEET=UNTZ_STROKE=none_FILL=(0,0,1,0.15);##!pragma_POLYGON(0,0,1,0.15)_121_132_110;####!_POLY_SHEET=UNTZ_STROKE=none_FILL=(0,0,1,0.15);##!pragma_POLYGON(0,0,1,0.15)_113_124_135_123;####!_POLY_SHEET=UNTZ_STROKE=none_FILL=(0,0,1,0.15);##!pragma_POLYGON(0,0,1,0.15)_114_125_136_124;####!_POLY_SHEET=UNTZ_STROKE=none_FILL=(0,0,1,0.15);##!pragma_POLYGON(0,0,1,0.15)_115_126_137_125;####!_POLY_SHEET=UNTZ_STROKE=none_FILL=(0,0,1,0.15);##!pragma_POLYGON(0,0,1,0.15)_116_127_138_126;####!_POLY_SHEET=UNTZ_STROKE=none_FILL=(0,0,1,0.15);##!pragma_POLYGON(0,0,1,0.15)_117_128_139_127;####!_POLY_SHEET=UNTZ_STROKE=none_FILL=(0,0,1,0.15);##!pragma_POLYGON(0,0,1,0.15)_118_129_140_128;####!_POLY_SHEET=UNTZ_STROKE=none_FILL=(0,0,1,0.15);##!pragma_POLYGON(0,0,1,0.15)_119_130_141_129;####!_POLY_SHEET=UNTZ_STROKE=none_FILL=(0,0,1,0.15);##!pragma_POLYGON(0,0,1,0.15)_120_131_142_130;####!_POLY_SHEET=UNTZ_STROKE=none_FILL=(0,0,1,0.15);##!pragma_POLYGON(0,0,1,0.15)_143_154_132;####!_POLY_SHEET=UNTZ_STROKE=none_FILL=(0,0,1,0.15);##!pragma_POLYGON(0,0,1,0.15)_134_145_156_144;####!_POLY_SHEET=UNTZ_STROKE=none_FILL=(0,0,1,0.15);##!pragma_POLYGON(0,0,1,0.15)_135_146_157_145;####!_POLY_SHEET=UNTZ_STROKE=none_FILL=(0,0,1,0.15);##!pragma_POLYGON(0,0,1,0.15)_136_147_158_146;####!_POLY_SHEET=UNTZ_STROKE=none_FILL=(0,0,1,0.15);##!pragma_POLYGON(0,0,1,0.15)_137_148_159_147;####!_POLY_SHEET=UNTZ_STROKE=none_FILL=(0,0,1,0.15);##!pragma_POLYGON(0,0,1,0.15)_138_149_160_148;####!_POLY_SHEET=UNTZ_STROKE=none_FILL=(0,0,1,0.15);##!pragma_POLYGON(0,0,1,0.15)_139_150_161_149;####!_POLY_SHEET=UNTZ_STROKE=none_FILL=(0,0,1,0.15);##!pragma_POLYGON(0,0,1,0.15)_140_151_162_150;####!_POLY_SHEET=UNTZ_STROKE=none_FILL=(0,0,1,0.15);##!pragma_POLYGON(0,0,1,0.15)_141_152_163_151;####!_POLY_SHEET=UNTZ_STROKE=none_FILL=(0,0,1,0.15);##!pragma_POLYGON(0,0,1,0.15)_142_153_164_152;####!_POLY_SHEET=UNTZ_STROKE=none_FILL=(0,0,1,0.15);##!pragma_POLYGON(0,0,1,0.15)_165_176_154;####!_POLY_SHEET=UNTZ_STROKE=none_FILL=(0,0,1,0.15);##!pragma_POLYGON(0,0,1,0.15)_155_166_177_165;####!_POLY_SHEET=UNTZ_STROKE=none_FILL=(0,0,1,0.15);##!pragma_POLYGON(0,0,1,0.15)_156_167_178_166;####!_POLY_SHEET=UNTZ_STROKE=none_FILL=(0,0,1,0.15);##!pragma_POLYGON(0,0,1,0.15)_157_168_179_167;####!_POLY_SHEET=UNTZ_STROKE=none_FILL=(0,0,1,0.15);##!pragma_POLYGON(0,0,1,0.15)_158_169_180_168;####!_POLY_SHEET=UNTZ_STROKE=none_FILL=(0,0,1,0.15);##!pragma_POLYGON(0,0,1,0.15)_159_170_181_169;####!_POLY_SHEET=UNTZ_STROKE=none_FILL=(0,0,1,0.15);##!pragma_POLYGON(0,0,1,0.15)_160_171_182_170;####!_POLY_SHEET=UNTZ_STROKE=none_FILL=(0,0,1,0.15);##!pragma_POLYGON(0,0,1,0.15)_161_172_183_171;####!_POLY_SHEET=UNTZ_STROKE=none_FILL=(0,0,1,0.15);##!pragma_POLYGON(0,0,1,0.15)_162_173_184_172;####!_POLY_SHEET=UNTZ_STROKE=none_FILL=(0,0,1,0.15);##!pragma_POLYGON(0,0,1,0.15)_163_174_185_173;####!_POLY_SHEET=UNTZ_STROKE=none_FILL=(0,0,1,0.15);##!pragma_POLYGON(0,0,1,0.15)_164_175_186_174;####!_POLY_SHEET=UNTZ_STROKE=none_FILL=(0,0,1,0.15);##!pragma_POLYGON(0,0,1,0.15)_187_198_176;####!_POLY_SHEET=UNTZ_STROKE=none_FILL=(0,0,1,0.15);##!pragma_POLYGON(0,0,1,0.15)_177_188_199_187;####!_POLY_SHEET=UNTZ_STROKE=none_FILL=(0,0,1,0.15);##!pragma_POLYGON(0,0,1,0.15)_178_189_200_188;####!_POLY_SHEET=UNTZ_STROKE=none_FILL=(0,0,1,0.15);##!pragma_POLYGON(0,0,1,0.15)_179_190_201_189;####!_POLY_SHEET=UNTZ_STROKE=none_FILL=(0,0,1,0.15);##!pragma_POLYGON(0,0,1,0.15)_180_191_202_190;####!_POLY_SHEET=UNTZ_STROKE=none_FILL=(0,0,1,0.15);##!pragma_POLYGON(0,0,1,0.15)_181_192_203_191;####!_POLY_SHEET=UNTZ_STROKE=none_FILL=(0,0,1,0.15);##!pragma_POLYGON(0,0,1,0.15)_182_193_204_192;####!_POLY_SHEET=UNTZ_STROKE=none_FILL=(0,0,1,0.15);##!pragma_POLYGON(0,0,1,0.15)_183_194_205_193;####!_POLY_SHEET=UNTZ_STROKE=none_FILL=(0,0,1,0.15);##!pragma_POLYGON(0,0,1,0.15)_184_195_206_194;####!_POLY_SHEET=UNTZ_STROKE=none_FILL=(0,0,1,0.15);##!pragma_POLYGON(0,0,1,0.15)_185_196_207_195;####!_POLY_SHEET=UNTZ_STROKE=none_FILL=(0,0,1,0.15);##!pragma_POLYGON(0,0,1,0.15)_186_197_208_196;####!_POLY_SHEET=UNTZ_STROKE=none_FILL=(0,0,1,0.15);##!pragma_POLYGON(0,0,1,0.15)_209_220_198;####!_POLY_SHEET=UNTZ_STROKE=none_FILL=(0,0,1,0.15);##!pragma_POLYGON(0,0,1,0.15)_199_210_221_209;####!_POLY_SHEET=UNTZ_STROKE=none_FILL=(0,0,1,0.15);##!pragma_POLYGON(0,0,1,0.15)_200_211_222_210;####!_POLY_SHEET=UNTZ_STROKE=none_FILL=(0,0,1,0.15);##!pragma_POLYGON(0,0,1,0.15)_201_212_223_211;####!_POLY_SHEET=UNTZ_STROKE=none_FILL=(0,0,1,0.15);##!pragma_POLYGON(0,0,1,0.15)_202_213_224_212;####!_POLY_SHEET=UNTZ_STROKE=none_FILL=(0,0,1,0.15);##!pragma_POLYGON(0,0,1,0.15)_203_214_225_213;####!_POLY_SHEET=UNTZ_STROKE=none_FILL=(0,0,1,0.15);##!pragma_POLYGON(0,0,1,0.15)_204_215_226_214;####!_POLY_SHEET=UNTZ_STROKE=none_FILL=(0,0,1,0.15);##!pragma_POLYGON(0,0,1,0.15)_205_216_227_215;####!_POLY_SHEET=UNTZ_STROKE=none_FILL=(0,0,1,0.15);##!pragma_POLYGON(0,0,1,0.15)_206_217_228_216;####!_POLY_SHEET=UNTZ_STROKE=none_FILL=(0,0,1,0.15);##!pragma_POLYGON(0,0,1,0.15)_207_218_229_217;####!_POLY_SHEET=UNTZ_STROKE=none_FILL=(0,0,1,0.15);##!pragma_POLYGON(0,0,1,0.15)_208_219_230_218;####!_POLY_SHEET=UNTZ_STROKE=none_FILL=(0,0,1,0.15);##!pragma_POLYGON(0,0,1,0.15)_21;####!_POLY_SHEET=UNTZ_STROKE=none_FILL=(0,0,1,0.15);##!pragma_POLYGON(0,0,1,0.15)_43;####!_POLY_SHEET=UNTZ_STROKE=none_FILL=(0,0,1,0.15);##!pragma_POLYGON(0,0,1,0.15)_65;####!_POLY_SHEET=UNTZ_STROKE=none_FILL=(0,0,1,0.15);##!pragma_POLYGON(0,0,1,0.15)_87;####!_POLY_SHEET=UNTZ_STROKE=none_FILL=(0,0,1,0.15);##!pragma_POLYGON(0,0,1,0.15)_109;####!_POLY_SHEET=UNTZ_STROKE=none_FILL=(0,0,1,0.15);##!pragma_POLYGON(0,0,1,0.15)_131;####!_POLY_SHEET=UNTZ_STROKE=none_FILL=(0,0,1,0.15);##!pragma_POLYGON(0,0,1,0.15)_153;####!_POLY_SHEET=UNTZ_STROKE=none_FILL=(0,0,1,0.15);##!pragma_POLYGON(0,0,1,0.15)_175;####!_POLY_SHEET=UNTZ_STROKE=none_FILL=(0,0,1,0.15);##!pragma_POLYGON(0,0,1,0.15)_197;####!_POLY_SHEET=UNTZ_STROKE=none_FILL=(0,0,1,0.15);##!pragma_POLYGON(0,0,1,0.15)_219;####!_POLY_SHEET=ANTIX_STROKE=none_FILL=(1,0,0,0.15);##!pragma_POLYGON(1,0,0,0.15)_40_62_51;####!_POLY_SHEET=ANTIX_STROKE=none_FILL=(1,0,0,0.15);##!pragma_POLYGON(1,0,0,0.15)_41;####!_POLY_SHEET=ANTIX_STROKE=none_FILL=(1,0,0,0.15);##!pragma_POLYGON(1,0,0,0.15)_53_63;####!_POLY_SHEET=ANTIX_STROKE=none_FILL=(1,0,0,0.15);##!pragma_POLYGON(1,0,0,0.15)_62_74_84_73;####!_POLY_SHEET=ANTIX_STROKE=none_FILL=(1,0,0,0.15);##!pragma_POLYGON(1,0,0,0.15)_64_75;####!_POLY_SHEET=ANTIX_STROKE=none_FILL=(1,0,0,0.15);##!pragma_POLYGON(1,0,0,0.15)_76_86;####!_POLY_SHEET=ANTIX_STROKE=none_FILL=(1,0,0,0.15);##!pragma_POLYGON(1,0,0,0.15)_85_97_107_96;####!_POLY_SHEET=ANTIX_STROKE=none_FILL=(1,0,0,0.15);##!pragma_POLYGON(1,0,0,0.15)_121_143_132;####!_POLY_SHEET=ANTIX_STROKE=none_FILL=(1,0,0,0.15);##!pragma_POLYGON(1,0,0,0.15)_122;####!_POLY_SHEET=ANTIX_STROKE=none_FILL=(1,0,0,0.15);##!pragma_POLYGON(1,0,0,0.15)_134_144;####!_POLY_SHEET=ANTIX_STROKE=none_FILL=(1,0,0,0.15);##!pragma_POLYGON(1,0,0,0.15)_143_155_165_154;####!_POLY_SHEET=ANTIZ_STROKE=none_FILL=(0,0,1,0.15);##!pragma_POLYGON(0,0,1,0.15)_30_41_40;####!_POLY_SHEET=ANTIZ_STROKE=none_FILL=(0,0,1,0.15);##!pragma_POLYGON(0,0,1,0.15)_31_42_53_41;####!_POLY_SHEET=ANTIZ_STROKE=none_FILL=(0,0,1,0.15);##!pragma_POLYGON(0,0,1,0.15)_63_74;####!_POLY_SHEET=ANTIZ_STROKE=none_FILL=(0,0,1,0.15);##!pragma_POLYGON(0,0,1,0.15)_62;####!_POLY_SHEET=ANTIZ_STROKE=none_FILL=(0,0,1,0.15);##!pragma_POLYGON(0,0,1,0.15)_54_65_76_64;####!_POLY_SHEET=ANTIZ_STROKE=none_FILL=(0,0,1,0.15);##!pragma_POLYGON(0,0,1,0.15)_86_97;####!_POLY_SHEET=ANTIZ_STROKE=none_FILL=(0,0,1,0.15);##!pragma_POLYGON(0,0,1,0.15)_75_85;####!_POLY_SHEET=ANTIZ_STROKE=none_FILL=(0,0,1,0.15);##!pragma_POLYGON(0,0,1,0.15)_111_122_121;####!_POLY_SHEET=ANTIZ_STROKE=none_FILL=(0,0,1,0.15);##!pragma_POLYGON(0,0,1,0.15)_112_123_134_122;####!_POLY_SHEET=ANTIZ_STROKE=none_FILL=(0,0,1,0.15);##!pragma_POLYGON(0,0,1,0.15)_144_155;####!_POLY_SHEET=ANTIZ_STROKE=none_FILL=(0,0,1,0.15);##!pragma_POLYGON(0,0,1,0.15)_143;####!_POLY_SHEET=PRODX_STROKE=none_FILL=(1,0,0,0.15);##!pragma_POLYGON(1,0,0,0.15)_75_64_76_86_97_107_96_85;####!_POLY_SHEET=PRODX_STROKE=none_FILL=(1,0,0,0.15);##!pragma_POLYGON(1,0,0,0.15)_41_53_63_74_84_73_51_40;####!_POLY_SHEET=PRODX_STROKE=none_FILL=(1,0,0,0.15);##!pragma_POLYGON(1,0,0,0.15)_122_134_144_155_165_154_132_121;####!_POLY_SHEET=PRODZ_STROKE=none_FILL=(0,0,1,0.15);##!pragma_POLYGON(0,0,1,0.15)_64_54_65_76_86_97_85_75;####!_POLY_SHEET=PRODZ_STROKE=none_FILL=(0,0,1,0.15);##!pragma_POLYGON(0,0,1,0.15)_31_42_53_63_74_62_40_30;####!_POLY_SHEET=PRODZ_STROKE=none_FILL=(0,0,1,0.15);##!pragma_POLYGON(0,0,1,0.15)_112_123_134_144_155_143_121_111;####!_POLY_SHEET=GAUGEX_STROKE=none_FILL=(1,0,0,0.15);##!pragma_POLYGON(1,0,0,0.15)_40_62_51;####!_POLY_SHEET=GAUGEX_STROKE=none_FILL=(1,0,0,0.15);##!pragma_POLYGON(1,0,0,0.15)_41_62_51_40;####!_POLY_SHEET=GAUGEX_STROKE=none_FILL=(1,0,0,0.15);##!pragma_POLYGON(1,0,0,0.15)_41_53_63_62_51_40;####!_POLY_SHEET=GAUGEX_STROKE=none_FILL=(1,0,0,0.15);##!pragma_POLYGON(1,0,0,0.15)_64_75;####!_POLY_SHEET=GAUGEX_STROKE=none_FILL=(1,0,0,0.15);##!pragma_POLYGON(1,0,0,0.15)_64_76_86_75;####!_POLY_SHEET=GAUGEX_STROKE=none_FILL=(1,0,0,0.15);##!pragma_POLYGON(1,0,0,0.15)_121_143_132;####!_POLY_SHEET=GAUGEX_STROKE=none_FILL=(1,0,0,0.15);##!pragma_POLYGON(1,0,0,0.15)_122_143_132_121;####!_POLY_SHEET=GAUGEX_STROKE=none_FILL=(1,0,0,0.15);##!pragma_POLYGON(1,0,0,0.15)_122_134_144_143_132_121;####!_POLY_SHEET=GAUGEZ_STROKE=none_FILL=(0,0,1,0.15);##!pragma_POLYGON(0,0,1,0.15)_30_41_40;####!_POLY_SHEET=GAUGEZ_STROKE=none_FILL=(0,0,1,0.15);##!pragma_POLYGON(0,0,1,0.15)_31_42_53_41;####!_POLY_SHEET=GAUGEZ_STROKE=none_FILL=(0,0,1,0.15);##!pragma_POLYGON(0,0,1,0.15)_63_74;####!_POLY_SHEET=GAUGEZ_STROKE=none_FILL=(0,0,1,0.15);##!pragma_POLYGON(0,0,1,0.15)_54_65_76_64;####!_POLY_SHEET=GAUGEZ_STROKE=none_FILL=(0,0,1,0.15);##!pragma_POLYGON(0,0,1,0.15)_86_97;####!_POLY_SHEET=GAUGEZ_STROKE=none_FILL=(0,0,1,0.15);##!pragma_POLYGON(0,0,1,0.15)_111_122_121;####!_POLY_SHEET=GAUGEZ_STROKE=none_FILL=(0,0,1,0.15);##!pragma_POLYGON(0,0,1,0.15)_112_123_134_122;####!_POLY_SHEET=GAUGEZ_STROKE=none_FILL=(0,0,1,0.15);##!pragma_POLYGON(0,0,1,0.15)_144_155;####!_HIGHLIGHT_TARGET=QUBIT_QUBITS=52,133_COLOR=red;TICK;CX_19_9_91_80_68_57_183_172_160_149_42_32_107_97_167_157_144_134_236_226_16_6_81_71_98_87_70_59_173_163_135_124_227_216_66_55_158_147_223_212_146_136_72_61_211_201_164_153_148_138_132_121_79_69_96_85_56_46_73_62_229_218_171_161_225_214_221_210_198_187_209_199_47_36_35_25_162_151_139_128_150_140_127_117_86_76_75_64_123_113_215_205_177_166_60_50_22_11_49_38_152_142_114_103_206_195_45_34_110_99_33_23_137_126_202_191_125_115_51_40_190_180_84_74_217_207_213_203_116_105_58_48_208_197_112_101_89_78_204_193_63_53_200_189_188_178_26_15_165_155_14_4_31_20_141_130_129_119_37_27_54_43_102_92_194_184_156_145_39_29_179_168_28_17_93_82_185_174_24_13_12_2_77_67_181_170_104_94_169_159_196_186_100_90_192_182_154_143_18_8;TICK;CX_150_139_127_116_36_25_128_117_193_182_59_48_124_113_101_90_216_205_37_26_102_91_194_183_33_22_43_32_125_114_190_179_39_28_217_206_34_23_168_157_191_180_40_29_196_185_5_16_100_89_197_186_192_181_1_12_188_177_165_154_88_99_170_159_166_155_176_187_15_4_42_31_107_96_38_27_103_92_3_14_80_69_236_225_130_119_195_184_172_161_132_143_81_70_173_162_77_66_210_199_104_93_169_158_13_2_151_140_0_11_105_94_82_71_147_136_78_67_79_68_56_45_171_160_20_9_7_18_167_156_57_46_44_55_149_138_214_203_8_19_85_74_145_134_110_121_123_112_17_6_215_204_64_53_174_163_146_135_211_200_60_49_87_76_152_141_212_201_189_178_148_137_84_73_61_50_126_115_153_142_218_207_213_202_62_51_209_198_58_47_35_24;TICK;MX_1_3_5_7_8_33_35_37_39_41_42_56_58_60_62_63_75_77_79_81_84_86_100_102_104_107_122_123_125_127_132_144_146_148_150_152_165_167_169_171_173_188_190_192_194_196_209_211_213_215_217_231_232_233_234_235_236_237_238_239_240;TICK;M_2_4_6_9_11_21_23_25_27_29_32_46_48_50_53_55_65_67_69_71_74_76_90_92_94_99_109_113_115_117_119_121_131_134_136_138_140_142_143_155_157_159_161_163_175_178_180_182_184_186_187_199_201_203_205_207_219;TICK;R_2_4_6_9_11_21_23_25_27_29_32_46_48_50_53_55_65_67_69_71_74_76_90_92_94_99_109_113_115_117_119_121_131_134_136_138_140_142_143_155_157_159_161_163_175_178_180_182_184_186_187_199_201_203_205_207_219;RX_1_3_5_7_8_33_35_37_39_41_42_56_58_60_62_63_75_77_79_81_84_86_100_102_104_107_122_123_125_127_132_144_146_148_150_152_165_167_169_171_173_188_190_192_194_196_209_211_213_215_217_231_232_233_234_235_236_237_238_239_240;TICK;CX_150_139_127_116_36_25_128_117_193_182_59_48_124_113_101_90_216_205_37_26_102_91_194_183_33_22_43_32_125_114_190_179_39_28_217_206_34_23_168_157_191_180_40_29_196_185_5_16_100_89_197_186_192_181_1_12_188_177_165_154_88_99_170_159_166_155_176_187_15_4_42_31_107_96_38_27_103_92_3_14_80_69_236_225_130_119_195_184_172_161_132_143_81_70_173_162_77_66_210_199_104_93_169_158_13_2_151_140_0_11_105_94_82_71_147_136_78_67_79_68_56_45_171_160_20_9_7_18_167_156_57_46_44_55_149_138_214_203_8_19_85_74_145_134_110_121_123_112_17_6_215_204_64_53_174_163_146_135_211_200_60_49_87_76_152_141_212_201_189_178_148_137_84_73_61_50_126_115_153_142_218_207_213_202_62_51_209_198_58_47_35_24;TICK;CX_19_9_91_80_68_57_183_172_160_149_42_32_107_97_167_157_144_134_236_226_16_6_81_71_98_87_70_59_173_163_135_124_227_216_66_55_158_147_223_212_146_136_72_61_211_201_164_153_148_138_132_121_79_69_96_85_56_46_73_62_229_218_171_161_225_214_221_210_198_187_209_199_47_36_35_25_162_151_139_128_150_140_127_117_86_76_75_64_123_113_215_205_177_166_60_50_22_11_49_38_152_142_114_103_206_195_45_34_110_99_33_23_137_126_202_191_125_115_51_40_190_180_84_74_217_207_213_203_116_105_58_48_208_197_112_101_89_78_204_193_63_53_200_189_188_178_26_15_165_155_14_4_31_20_141_130_129_119_37_27_54_43_102_92_194_184_156_145_39_29_179_168_28_17_93_82_185_174_24_13_12_2_77_67_181_170_104_94_169_159_196_186_100_90_192_182_154_143_18_8;TICK;CX_95_84_19_9_91_80_68_57_183_172_160_149_167_157_85_75_232_222_16_6_81_71_108_98_70_59_173_163_135_124_17_7_227_216_66_55_76_65_158_147_223_212_146_136_72_61_211_201_234_224_83_73_148_138_240_230_79_69_56_46_229_218_171_161_20_10_225_214_74_63_122_112_221_210_198_187_209_199_47_36_35_25_162_151_139_128_150_140_134_123_127_117_238_228_215_205_177_166_64_54_118_107_60_50_22_11_49_38_152_142_114_103_206_195_45_34_110_99_33_23_137_126_202_191_125_115_190_180_217_207_121_111_213_203_116_105_58_48_40_30_89_78_204_193_53_42_200_189_188_178_26_15_14_4_141_130_129_119_106_96_37_27_102_92_194_184_156_145_97_86_179_168_93_82_120_109_185_174_24_13_12_2_29_18_77_67_181_170_104_94_176_165_169_159_196_186_100_90_192_182_41_31;TICK;CX_150_139_127_116_36_25_128_117_193_182_59_48_216_205_65_54_37_26_102_91_194_183_33_22_125_114_190_179_217_206_34_23_168_157_191_180_40_29_196_185_5_16_100_89_155_144_192_181_1_12_41_30_18_7_188_177_88_99_170_159_129_118_106_95_176_187_15_4_42_31_107_96_38_27_154_165_103_92_3_14_80_69_130_119_195_184_172_161_232_221_81_70_108_97_173_162_77_66_210_199_104_93_169_158_109_98_13_2_151_140_0_11_105_94_82_71_147_136_6_17_78_67_234_223_83_72_240_229_79_68_56_45_171_160_167_156_57_46_44_55_149_138_214_203_8_19_85_74_123_112_238_227_215_204_64_53_9_20_174_163_146_135_211_200_60_49_152_141_212_201_189_178_148_137_84_73_61_50_126_115_121_110_218_207_213_202_122_111_209_198_58_47_35_24;TICK;M_2_4_7_11_21_23_25_27_30_31_43_46_48_50_54_55_62_63_67_69_71_73_75_86_87_92_94_96_98_99_111_112_115_117_119_131_136_138_140_143_144_153_157_159_161_163_165_175_178_180_182_184_187_197_199_201_203_205_207_219;MX_1_3_5_6_8_9_33_35_37_40_56_58_60_64_77_79_81_83_85_100_102_104_106_108_121_125_127_129_146_148_150_152_167_169_171_173_188_190_192_194_196_209_211_213_215_217_231_232_233_234_235_236_237_238_239_240;TICK;R_2_4_7_11_21_23_25_27_30_31_43_46_48_50_54_55_62_63_67_69_71_73_75_86_87_92_94_96_98_99_111_112_115_117_119_131_136_138_140_143_144_153_157_159_161_163_165_175_178_180_182_184_187_197_199_201_203_205_207_219;RX_1_3_5_6_8_9_33_35_37_40_56_58_60_64_77_79_81_83_85_100_102_104_106_108_121_125_127_129_146_148_150_152_167_169_171_173_188_190_192_194_196_209_211_213_215_217_231_232_233_234_235_236_237_238_239_240;TICK;CX_150_139_127_116_36_25_128_117_193_182_59_48_216_205_65_54_37_26_102_91_194_183_33_22_125_114_190_179_217_206_34_23_168_157_191_180_40_29_196_185_5_16_100_89_155_144_192_181_1_12_41_30_18_7_188_177_88_99_170_159_129_118_106_95_176_187_15_4_42_31_107_96_38_27_154_165_103_92_3_14_80_69_130_119_195_184_172_161_232_221_81_70_108_97_173_162_77_66_210_199_104_93_169_158_109_98_13_2_151_140_0_11_105_94_82_71_147_136_6_17_78_67_234_223_83_72_240_229_79_68_56_45_171_160_167_156_57_46_44_55_149_138_214_203_8_19_85_74_123_112_238_227_215_204_64_53_9_20_174_163_146_135_211_200_60_49_152_141_212_201_189_178_148_137_84_73_61_50_126_115_121_110_218_207_213_202_122_111_209_198_58_47_35_24;TICK;CX_95_84_19_9_91_80_68_57_183_172_160_149_167_157_85_75_232_222_16_6_81_71_108_98_70_59_173_163_135_124_17_7_227_216_66_55_76_65_158_147_223_212_146_136_72_61_211_201_234_224_83_73_148_138_240_230_79_69_56_46_229_218_171_161_20_10_225_214_74_63_122_112_221_210_198_187_209_199_47_36_35_25_162_151_139_128_150_140_134_123_127_117_238_228_215_205_177_166_64_54_118_107_60_50_22_11_49_38_152_142_114_103_206_195_45_34_110_99_33_23_137_126_202_191_125_115_190_180_217_207_121_111_213_203_116_105_58_48_40_30_89_78_204_193_53_42_200_189_188_178_26_15_14_4_141_130_129_119_106_96_37_27_102_92_194_184_156_145_97_86_179_168_93_82_120_109_185_174_24_13_12_2_29_18_77_67_181_170_104_94_176_165_169_159_196_186_100_90_192_182_41_31;TICK;CX_78_68_170_160_235_225_224_213_166_156_231_221_15_5_32_21_220_209_107_97_69_58_161_150_92_81_157_146_184_173_237_227_233_223_17_7_210_200_174_164_13_3_151_141_55_45_105_95_67_56_82_72_44_33_147_137_159_148_239_229_155_144_132_121_96_85_73_62_61_51_20_10_218_208_57_47_122_112_149_139_214_204_203_192_145_135_134_123_199_188_48_37_75_64_140_129_71_60_136_125_59_49_124_114_163_152_101_91_228_217_216_206_212_202_189_179_201_190_142_131_84_74_46_35_23_12_126_116_138_127_115_104_230_219_111_100_99_89_226_215_191_181_187_177_36_26_30_19_63_53_25_14_128_118_90_79_193_183_182_171_113_102_178_167_27_16_205_194_119_108_50_39_38_28_103_93_80_70_207_196_130_120_195_185_34_24_172_162_11_1_88_77_168_158_180_169_222_211_176_165_186_175_117_106_94_83_154_143_41_31_18_8;TICK;CX_150_139_127_116_187_176_36_25_128_117_193_182_59_48_124_113_101_90_216_205_37_26_102_91_194_183_125_114_190_179_39_28_217_206_34_23_131_120_168_157_99_88_191_180_196_185_100_89_192_181_41_30_188_177_14_3_170_159_129_118_106_95_166_155_231_220_66_77_2_13_107_96_38_27_154_165_103_92_80_69_130_119_195_184_172_161_16_5_81_70_108_97_173_162_12_1_210_199_104_93_4_15_169_158_151_140_0_11_55_44_105_94_82_71_198_209_147_136_6_17_239_228_78_67_143_132_83_72_235_224_19_8_175_164_79_68_56_45_171_160_7_18_57_46_167_156_149_138_214_203_85_74_145_134_110_121_237_226_21_10_233_222_215_204_9_20_174_163_64_53_146_135_211_200_60_49_152_141_212_201_189_178_148_137_84_73_61_50_126_115_218_207_122_111_213_202_22_33_62_51_76_86_219_208_58_47_35_24;TICK;M_1_3_5_8_10_24_26_28_33_43_45_47_49_51_53_65_68_70_72_74_77_87_89_91_93_95_97_109_114_116_118_120_121_132_135_137_139_141_153_156_158_160_162_164_165_177_179_181_183_185_197_200_202_204_206_208_209;MX_0_2_4_6_7_9_34_36_38_41_55_57_59_61_63_75_76_78_80_82_84_99_101_103_105_107_122_124_126_128_130_143_145_147_149_151_166_168_170_172_174_187_189_191_193_195_210_212_214_216_218_231_232_233_234_235_236_237_238_239_240;TICK;R_1_3_5_8_10_24_26_28_33_43_45_47_49_51_53_65_68_70_72_74_77_87_89_91_93_95_97_109_114_116_118_120_121_132_135_137_139_141_153_156_158_160_162_164_165_177_179_181_183_185_197_200_202_204_206_208_209;RX_0_2_4_6_7_9_34_36_38_41_55_57_59_61_63_75_76_78_80_82_84_99_101_103_105_107_122_124_126_128_130_143_145_147_149_151_166_168_170_172_174_187_189_191_193_195_210_212_214_216_218_231_232_233_234_235_236_237_238_239_240;TICK;CX_150_139_127_116_187_176_36_25_128_117_193_182_59_48_124_113_101_90_216_205_37_26_102_91_194_183_125_114_190_179_39_28_217_206_34_23_131_120_168_157_99_88_191_180_196_185_100_89_192_181_41_30_188_177_14_3_170_159_129_118_106_95_166_155_231_220_66_77_2_13_107_96_38_27_154_165_103_92_80_69_130_119_195_184_172_161_16_5_81_70_108_97_173_162_12_1_210_199_104_93_4_15_169_158_151_140_0_11_55_44_105_94_82_71_198_209_147_136_6_17_239_228_78_67_143_132_83_72_235_224_19_8_175_164_79_68_56_45_171_160_7_18_57_46_167_156_149_138_214_203_85_74_145_134_110_121_237_226_21_10_233_222_215_204_9_20_174_163_64_53_146_135_211_200_60_49_152_141_212_201_189_178_148_137_84_73_61_50_126_115_218_207_122_111_213_202_22_33_62_51_76_86_219_208_58_47_35_24;TICK;CX_78_68_170_160_235_225_224_213_166_156_231_221_15_5_32_21_220_209_107_97_69_58_161_150_92_81_157_146_184_173_237_227_233_223_17_7_210_200_174_164_13_3_151_141_55_45_105_95_67_56_82_72_44_33_147_137_159_148_239_229_155_144_132_121_96_85_73_62_61_51_20_10_218_208_57_47_122_112_149_139_214_204_203_192_145_135_134_123_199_188_48_37_75_64_140_129_71_60_136_125_59_49_124_114_163_152_101_91_228_217_216_206_212_202_189_179_201_190_142_131_84_74_46_35_23_12_126_116_138_127_115_104_230_219_111_100_99_89_226_215_191_181_187_177_36_26_30_19_63_53_25_14_128_118_90_79_193_183_182_171_113_102_178_167_27_16_205_194_119_108_50_39_38_28_103_93_80_70_207_196_130_120_195_185_34_24_172_162_11_1_88_77_168_158_180_169_222_211_176_165_186_175_117_106_94_83_154_143_41_31_18_8}
\maketitle

\begin{abstract}
    Utility-scale solid-state quantum devices will need to fabricate quantum devices at scale using imperfect processes.
    By introducing tolerance to fabrication defects into the design of the quantum devices, we can improve the yield of usable quantum chips and lower the cost of useful systems.
    Automated Compilation Including Dropouts (ACID) is a framework that works in the ancilla-free (or `middle-out') paradigm, to generate syndrome extraction circuits for general stabiliser codes in the presence of defective couplers or qubits. 
    In the ancilla-free paradigm, we do not designate particular qubits as measurement ancillas, instead measuring stabilisers using any of the data qubits in their support.
    This approach leads to a great deal of flexibility in how syndrome extraction circuits can be implemented.
    Given a stabiliser code and information about the functional and defective couplers between the qubits, ACID works by constructing and solving an optimisation problem within the ancilla-free paradigm to find a short syndrome extraction circuit.
    Applied to the surface code, ACID produces syndrome extraction circuits of depth between $1\times$ (no overhead) and $1.5\times$ relative to the depth of defect-free circuits. The original LUCI algorithm, the best prior art, yielded a $2 \times$ overhead, so ACID offers a significant time saving.
    The yield of surface code chips with a logical error rate at most $10\times$ the dropout-free baseline is up to $3\times$ higher using ACID than using original LUCI. 
    I demonstrate the broad applicability of ACID by compiling syndrome extraction circuits for bivariate bicycle codes and the colour code. 
    For these circuits, we incur a circuit-depth overhead of between $1\times$ (no overhead) and $2.5\times$ relative to defect-free circuits.
    Depending on the underlying connectivity, we can often accommodate isolated defective couplers and qubits with only an order-unity impact on the logical error rate.
    I believe this work is the first to simulate both of these families of codes in the presence of fabrication defects.

\end{abstract}

\section{Introduction}

To solve problems of broad scientific and commercial interest, quantum computers will need to maintain the integrity of the stored logical information for a duration billions or trillions of times longer than the decoherence time of their most fundamental components.
In particular, for the artificial qubits (superconducting and spin) which are the main target of this paper, fabrication and control techniques must be improved to bring down the decoherence probabilities associated with each operation.
To help bridge the gap between current physical error rates and the logical error rates we need, we can exponentially amplify the benefit of such engineering advances at the cost of introducing many additional physical qubits, by encoding our information redundantly in a quantum error-correcting code.

Since these quantum error-correcting codes were created to handle imperfections in circuit execution, it would seem natural that they could also handle known, permanent imperfections in the quantum devices themselves.
Even state-of-the-art fabrication techniques do not have a perfect yield \cite{aruteQuantumSupremacyUsing2019, acharyaQuantumErrorCorrection2025b}, and requiring that each quantum chip is perfect---that every qubit and every coupler between the qubits is functional and high fidelity---would incur a substantial, perhaps unviable, manufacturing overhead in post-selecting for chips that turned out perfectly.
I note that the classical semiconductor industry, whose products are far less delicate devices than transmons or SETs, routinely builds redundancy into its chip architectures to accommodate inevitable defects.
In the same way that QECCs amplify improvements in gate execution, some level of defect-tolerance in the architecture of QPUs could substantially amplify the gain in device yield afforded by improvements in fabrication.

 I work with the assumption that certain qubits and couplers on our device, known at the time of circuit compilation, are completely non-functional.
In the case of broken (`dropped') couplers, this means that the two qubits the coupler connects can no longer interact with one another, whereas `dropped' qubits cannot interact with any of their neighbours; dropping a qubit is equivalent to dropping all its connected couplers.
It seems likely similar techniques might be useful where a qubit is functional but is known to be highly error-prone.

There is plenty of literature proposing techniques to handle dropouts (as I will  call them henceforth) in the surface code \cite{augerFaulttoleranceThresholdsSurface2017,siegelAdaptiveSurfaceCode2023,strikisQuantumComputingScalable2023, linCodesignQuantumErrorcorrecting2024, lerouxSnakesLaddersAdapting2025} (see also \cite{mclauchlanAccommodatingFabricationDefects2024}, where dropouts on Floquet codes are explored).
The state-of-the-art is the LUCI framework \cite{debroyLUCISurfaceCode2024, higgottHandlingFabricationDefects2025}, which can handle multiple dropped couplers and qubits, often without losing code distance.
However, LUCI syndrome extraction circuits have a circuit depth twice that in the dropout-free case.
This results in an overall increase in computation time by at least the same factor, and in addition may increase logical error rates as each round provides a longer window for errors to accrue. 

In this work, I introduce Automated Compilation Including Dropouts (ACID), a framework within which any CSS stabiliser code can be compiled to a syndrome extraction cycle that accommodates any pattern of dropouts, with any suitable underlying qubit connectivity.
ACID generalises LUCI, treating general codes in the ancilla-free  (or `middle-out') paradigm.
It then encodes the problem of finding a syndrome extraction circuit into an integer linear program (ILP) and optimises using standard techniques to find a short syndrome extraction circuit.
I demonstrate compilation of the surface code, the bivariate bicycle codes and the colour code, each using multiple underlying qubit connectivities.
A full characterisation of performance of the nine code-connectivity setups for up to three dropped qubits and three dropped couplers is given in Appendix \ref{sec:plots}.

I demonstrate ACID applied to the bivariate bicycle code assuming both the degree-5 connectivity introduced in \cite{shawLoweringConnectivityRequirements2025}, and an alternative degree-6 connectivity that introduces more redundancy and leads to greater robustness in qubit dropouts.
These result in syndrome extraction circuits of depth between $1\times$ and $2.5\times$ the dropout-free case, depending on the number of dropouts.
I applied ACID to both the `middle-out' and `superdense' formulations of the colour code presented in \cite{gidneyNewCircuitsOpen2023a}, trying the former on both a hex-grid and square-grid lattice and the latter on a square-grid lattice.
Again, the syndrome extraction circuits are of length between $1\times$ and $2.5\times$ the dropout-free circuits, with the three configurations each offering a different balance of qubit overhead, syndrome extraction circuit depth, and logical error rate.

\begin{figure}[H]
    \centering
    \includegraphics[width=0.8\linewidth]{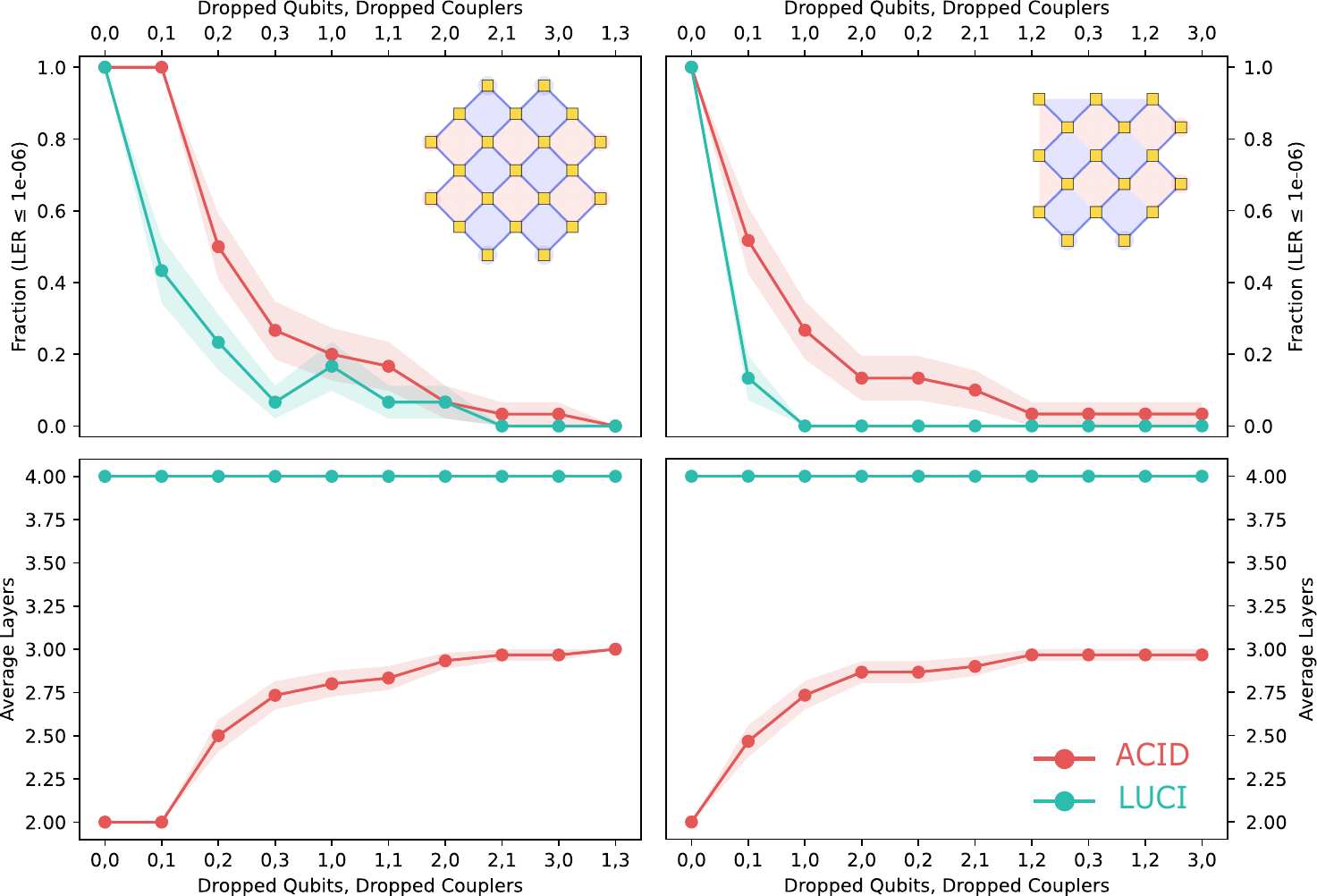}
    \caption[]{ 
    \textit{A comparison between ACID and the state-of-the-art LUCI algorithm \cite{debroyLUCISurfaceCode2024} when both are applied to the distance-11 surface code.}
    Top left: the proportion of cases where ACID and LUCI are able to produce syndrome extraction circuits with a logical error rate less than $10^{-6}$ for a distance-11 surface code implemented on a device with square-grid connectivity, assuming a random subset of qubits and couplers are defective.
    The numbers of defective qubits and defective couplers are shown on the $x$-axis, and the shaded region is the standard binomial error.
    Bottom left: In the same situations, the average number of `contraction layers' (Section \ref{sec:ancilla-free}) required to construct a round of syndrome extraction, plotted with the standard error in the mean.
    LUCI always requires 4 rounds, whereas ACID is observed to never require more than 3.
    Top right and bottom right: The same comparison for the distance-11 surface code implemented on a device with hex-grid connectivity.
    These plots are a summary of the full data provided in Appendix \ref{sec:plots}.}
    \label{fig:surfacesummaryplot}
\end{figure}

Applied to the surface code, ACID maintains code-capacity distance in the same situations that LUCI does, and can typically find syndrome extraction cycles that have either the same depth or $1.5\times$ the depth of the dropout-free circuits---a reduction in time-overhead of between $25\%$ and $50\%$.
Its distance-preserving properties are comparable: depending on whether we assume an underlying square-grid (degree-4) or hex-grid (degree-3) connectivity, we can tolerate the loss of certain qubits and couplers without losing distance.
I implemented a version of the LUCI algorithm as a special case of ACID with additional constraints (Appendix \ref{sec:luci-repro}), and find that ACID is able to maintain error rates within an order of magnitude of the dropout-free baseline up to $3\times$ more often than LUCI for small numbers of dropped qubits and couplers (Figure \ref{fig:surfacesummaryplot}).

The source code for ACID, with examples, is \href{https://github.com/riverlane/ACID}{available on GitHub \cite{RiverlaneACID}.}
The repository also hosts visualisations of syndrome extraction circuits produced by the algorithm.

\textit{
Update (January 2026): the comparison made here is to a reproduction of LUCI insofar as it is detailed in the original two works \cite{debroyLUCISurfaceCode2024, higgottHandlingFabricationDefects2025}. Shortly after the release of this manuscript, Anker and Debroy \cite{ankerOptimizedMeasurementSchedules2025} showed by means of a direct comparison that ACID produces surface code syndrome extraction circuits with higher error rates than those produced by the closed-source implementation of LUCI upon which the original LUCI papers were based, explaining that that their implementation contains heuristic optimisations that are not presented in the original work.
Anker and Debroy also detail a new version of LUCI using very similar techniques to those in the present work, but equipped with a set of additional, explicitly described optimisation heuristics, meaning the most up-to-date version of the LUCI algorithm is both more performant and more reproducible than previous versions, and is also capable of matching ACID's time overhead. 
In my view, the newest version of LUCI should be considered the state-of-the-art for tolerating fabrication defects in the surface code.
Promisingly, the heuristics presented therein are compatible with the general framework of ACID, and future work should merge the two with the aim of producing similarly optimised syndrome extraction circuits for other stabiliser codes.
}

\begin{figure}
    \centering
    \includegraphics[width=\linewidth]{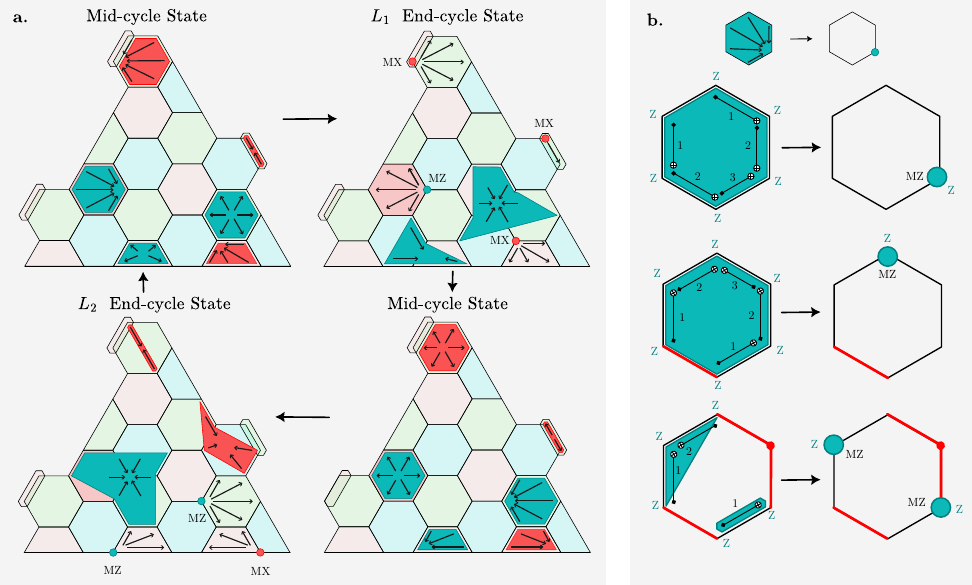}
    \caption[]{ 
    \textit{The ancilla-free approach to implementing quantum error-correcting codes.} \subfiglab{a.} A single round of one possible syndrome extraction cycle for the triangular colour code, that consists of two contraction layers.
    I've highlighted three $X$-type (red) and three $Z$-type (teal) stabilisers and track the corresponding instantaneous stabilisers through the syndrome extraction cycle.
    We start (top-left) the cycle in the mid-cycle state, which comes from the standard definition of the colour code, and contains no ancilla qubits. In the first round, the three $X$ stabilisers and one of the $Z$ stabilisers are contracted, and the remaining two $Z$ stabilisers are expanded.
    We perform three timesteps' worth of CNOT gates to take us to the second stage, known as the $L_1$ end-cycle state.
    The contracted stabilisers are now each localised onto a single qubit which we measure in the appropriate basis.
    We undo the CNOTs to return to the mid-cycle state.
    In the second layer, we contract a different subset of the stabilisers (producing the $L_2$ end-cycle state), such that each stabiliser is measured at least once over the two rounds. Note that the $X$ stabiliser in the bottom-right is measured in both layers, but is contracted onto a different qubit in each.

    {}\quad\subfiglab{ b.} 
    Here I show how a weight-six stabiliser with a cyclic local connectivity is affected by the presence of dropouts.
    In the top case, we have no dropouts and we arbitrarily choose a contraction schedule onto one of the qubits, subject only to compatibility with the schedules of neighbouring stabilisers.
    In the middle case, the presence of a dropped-out coupler forces us to use a different contraction schedule (note there are still two contraction schedules to choose from).
    In the bottom case, the presence of both a defective coupler and a defective qubit causes the remaining qubits in the stabiliser to become disconnected.
    In this case, we define `quasi-stabilisers' from the remaining connected components and define product stabilisers and gauge qubits from them, forming a subsystem code.
    We then measure these stabilisers separately, and classically multiply the outcomes of multiple quasi-stabilisers to form deterministic detectors.
}

\label{fig:intro-overview}
\end{figure}

\section{Ancilla-free QEC}\label{sec:ancilla-free}

Vasmer and Kubica introduced the `morphing' technique for stabiliser codes, in which a subset of the physical qubits in a stabiliser code are disentangled from the rest using a unitary circuit, leaving behind a different code that contains the same logical information but  has fewer physical qubits \cite{vasmerMorphingQuantumCodes2022} and different error-correction performance.
The ancilla-free (or `middle-out') view of QEC was first proposed in \cite{mcewenRelaxingHardwareRequirements2023a} for the surface code, generalised to the colour code in \cite{gidneyNewCircuitsOpen2023a} and generalised to the bivariate bicycle codes in \cite{shawLoweringConnectivityRequirements2025}.
All three works can be seen as an application of the original morphing technique.
In the ancilla-free framework, we map a Calderbank-Shor-Steane (CSS) stabiliser code with $n$ qubits onto $n$ physical qubits, without any additional qubits designated for measuring the stabilisers.
In other words, stabilisers are defined as subsets of the physical qubits along with a Pauli type ($X$/$Z$).

The ancilla-free framework in general requires the same number of physical qubits as in ancilla-based QEC to implement codes with the same distance and logical qubits, but has the key advantage that it allows far more flexibility in constructing syndrome extraction circuits.
In \cite{mcewenRelaxingHardwareRequirements2023a} and \cite{shawLoweringConnectivityRequirements2025} this flexibility is leveraged to reduce the connectivity required to implement surface and bivariate bicycle codes.
In the LUCI framework \cite{debroyLUCISurfaceCode2024, higgottHandlingFabricationDefects2025} and in the present work, this flexibility allows us to work around dropouts (and optionally maintain the reduced connectivity requirements).

One way to understand the ancilla-free paradigm is by considering the evolution of a set of privileged generators of the instantaneous stabiliser group (ISG) throughout a syndrome extraction round.
At the start of a round our qubits are stabilised by the stabilisers of the input code, and we take these as our privileged generators.
Confusingly, the input code is called the `mid-cycle code' and the qubits are said to be in the `mid-cycle state' at this point in the circuit.

To measure some stabiliser S of the mid-cycle code,  we can perform a  Clifford operation C, and then measure the single-qubit Pauli operator $P$, where  $C^{-1}SC = P$.
The process of localising the support of a stabiliser generator onto a single qubit is called `contracting'.
Furthermore, we can measure a subset of $m$ stabiliser elements in parallel if $\{S_1, S_2, ... S_m\}$ satisfy $C^{-1} S_j C = P_j$ for some set of single qubit Pauli operators $\{P_1, P_2, ... P_m\}$.  These Cliffords must also be broken down into CNOT gates that respect device connectivity.
We refer to these joint measurement operations as `contraction layers'.

These are depicted in Figure \ref{fig:intro-overview}a.
In the first half of each contraction layer, we choose some compatible subset of the stabilisers to measure, and perform CNOTs to perform the Clifford $C$.
This has the effect of contracting the support of each stabiliser so that it consists of a single qubit, which is called the `root qubit'.
This root qubit is then measured and reset in the appropriate basis.
The stabilisers that are not being contracted in a given round have their support modified in some complex way, according to $\bar{S} \rightarrow C^{-1}\bar{S}C$.
The form of these expanded stabilisers halfway through a syndrome extraction layer, given by $\{C^{-1}\bar{S_i}C\}$, defines the `end-cycle code', which contains the same logical information as the mid-cycle code, but uses fewer than $n$ qubits and in general has a reduced distance.
In the second half of each contraction layer, we perform the same CNOTs in reverse to enact $C^{-1}$, `un-contracting' the contracted stabilisers from their singular support on the root qubits and `un-expanding' the expanded stabilisers so all the stabiliser generators return to their original `mid-cycle' state.

In the next contraction layer, we choose a different (not necessarily disjoint) set of stabilisers to contract, and repeat the process of contracting these onto some chosen root qubits, measuring and resetting the root qubits, and undoing the contraction.
We perform as many of these layers as we need to measure each stabiliser at least once.
Note that in some cases, the ancilla-free and ancilla-based paradigms can describe the same syndrome extraction circuit: notably, in the most basic distance-$d$ rotated surface code circuits presented in \cite{mcewenRelaxingHardwareRequirements2023a}, the mid-cycle code has $\sim 2d^2$ physical qubits, and each end-cycle code can be viewed as an unrotated surface code supported on $\sim d^2$ physical `data' qubits alongside $\sim d^2$ disentangled `ancilla' qubits.
In this work, we privilege the mid-cycle code, fixing it and allowing the end-cycle codes to vary based on the connectivity and dropouts.

For the surface, colour and bivariate bicycle codes we consider, in the absence of dropouts, there are elegant constructions that allow us to measure all stabilisers in two layers, contracting and measuring half of the stabilisers in each.
In this work, the process of finding these contraction assignments is automated, so that ACID rediscovers the deliberate constructions of the existing literature in the absence of dropouts and finds new schedules in their presence.

As depicted in Figure \ref{fig:intro-overview}b, each stabiliser can usually be contracted and measured in multiple ways, and this redundancy can be exploited to measure stabilisers despite the presence of broken couplers.
There are cases (e.g. the bottom panel of Figure \ref{fig:intro-overview}b) where the presence of dropped couplers and/or qubits causes the support of stabilisers to be reduced, or even causes stabilisers to split into multiple parts.
When this happens, the objects produced may no longer commute with each other, or with the undamaged stabilisers.
In this case we deploy the theory of subsystem codes \cite{baconOperatorQuantumError2006}, allowing us to define product stabilisers made up of products of the anticommuting `quasi-stabilisers' that do commute with one another.

\section{Methods}

ACID takes as input an undirected graph $G$ whose nodes represent the qubits of our device, and whose edges represent the  couplers between them, assuming perfect fabrication.
I call this the `global connectivity graph' or just `the connectivity'.
It also takes a set of stabilisers defining the code, each of which is mapped to a subset of the nodes of the global connectivity graph.
The subgraph of $G$ induced by the nodes in the support of a given stabiliser is the `local connectivity graph' of that stabiliser.
Finally, we specify which qubits and couplers have dropped out.

\begin{figure}
    \centering
    \includegraphics[width=0.9\linewidth]{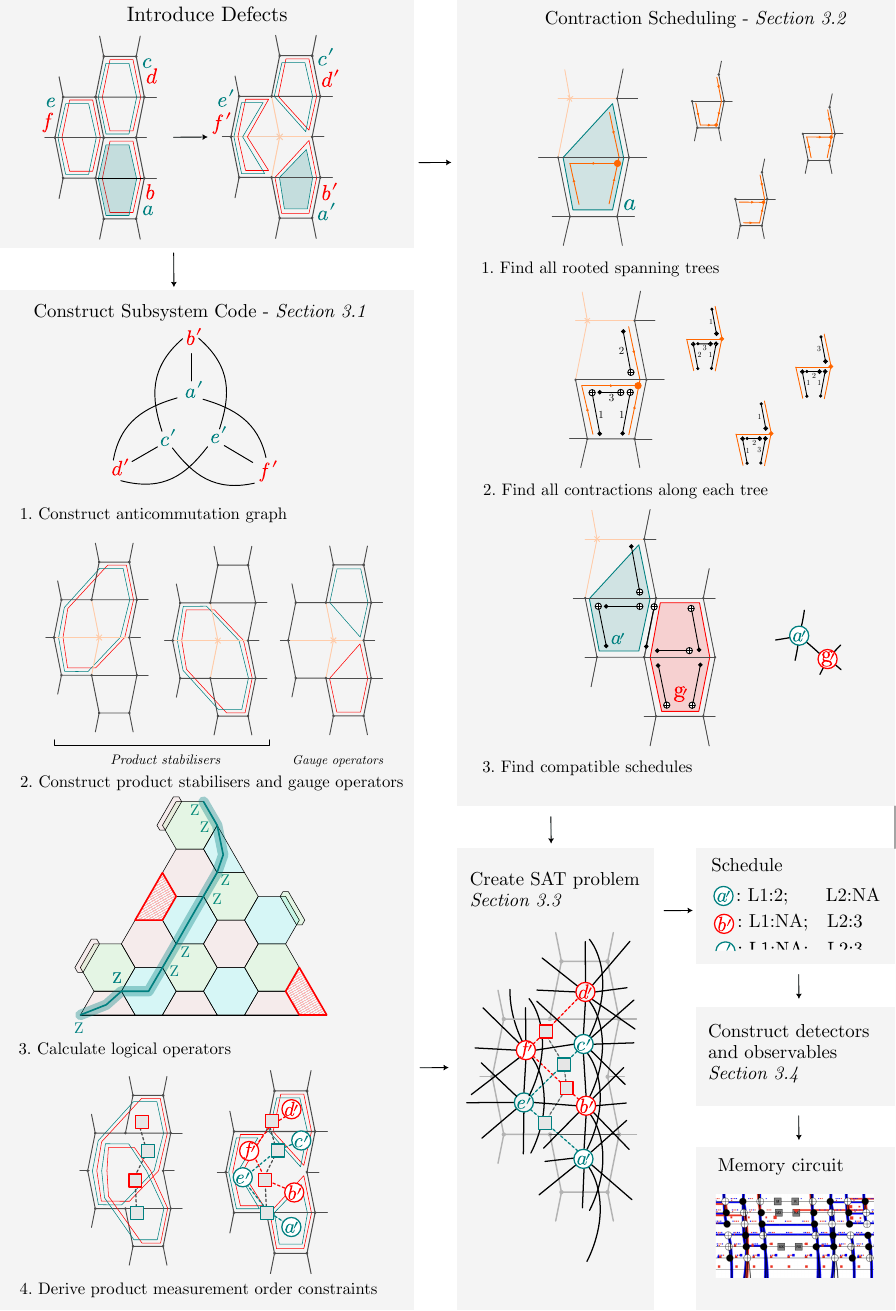}
    \caption{An overview of the process of compiling a memory circuit around defects.}
    \label{fig:overview}
\end{figure}

The end-to-end process of going from a stabiliser code, global connectivity graph, and set of dropouts, to a syndrome extraction circuit and ultimately to the memory experiments described in this work is summarised in Figure \ref{fig:overview}.

\subsection{Constructing a subsystem code from a defective stabiliser code}\label{sec:subsystem}

Each stabiliser is defined by its local connectivity graph.
The first step of ACID is to split each stabiliser into one or more `quasi-stabiliser' objects.
For each stabiliser, it removes any edges corresponding to dropped couplers or nodes corresponding to dropped qubits from the local connectivity graph, leaving a modified graph with one or more connected components (or zero components if every qubit has dropped out).
Each connected component defines a new operator, which I call a `quasi-stabiliser'.

In general, these quasi-stabilisers no longer all commute with one another.
We could simply choose not to measure these anticommuting quasi-stabilisers at all, but we would in doing so be throwing away useful information we can use to catch errors, and we will end up substantially reducing the distance of our code, as well as changing the number of logical qubits encoded.
We can in many cases maintain the code distance by packaging the anticommuting quasi-stabilisers into sets to form \textit{product stabilisers}, which do all commute with each other.
These product stabilisers are measured by physically contracting and measuring their constituent quasi-stabilisers (potentially over multiple syndrome extraction layers) and classically multiplying the results.

Define the quasi-stabiliser operators $Q$ as CSS-type Pauli strings corresponding to the physical sets of connected qubits from which we build our code.
I will in general notate elements of a set with lower-case letters, so that $Q = \{q_i\}$, and will sometimes interchangeably refer to operators and the physical constructions they describe, hopefully without ambiguity.
The quasi-stabilisers, which include the still-commuting untouched stabilisers whose commutation relations were not altered by the dropouts, form a (non-commuting) \textit{gauge group} that defines a subsystem code.
In order to operate the subsystem code---that is, to construct from it syndrome extraction circuits---it is sufficient to find a basis for the subsystem code as follows.

\begin{figure}    
    \centering
    \includegraphics[width=0.9\linewidth]{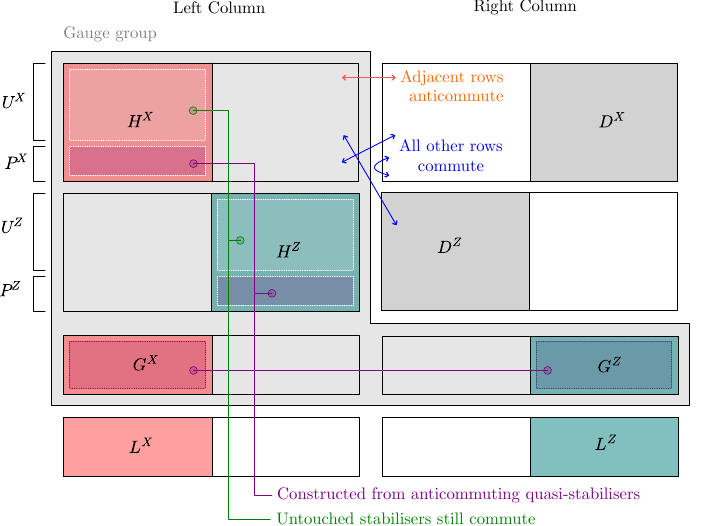}
    \caption{\textit{The commutation relations of a CSS subsystem code (symplectic basis).} 
    The left and right `columns' are each binary matrices, each of width $2n$ (twice the number of physical qubits) and height $n$, so that, taken together, the $2n$ rows form a complete basis for $F_2^{2n}$.
    Each row of either column represents a Pauli operator. Within each row the first $n$ bits represent the places where a Pauli operator contains an $X$, and the last $n$ bits represent where it contains a $Z$.
    Then the diagram can be read as a series of left-right pairs of Pauli operators, so that each operator anticommutes with its neighbour and commutes with all others.
    Within this representation, there are blocks corresponding to the parity check matrices $H^X$ and $H^Z$, the gauge operators $G^X$ and $G^Z$, and logical operators $L^X$ and $L^Z$.
    To complete the basis, I've included the destabiliser matrices $D^X$ and $D^Z$, although for practical purposes we do not need to compute these.
    The bulk of the stabilisers are the existing untouched stabilisers $U^X$ and $U^Z$, with the remaining stabilisers and gauge operators formed as products of the `touched' quasi-stabilisers.
    Viewed this way, it is clear that a subsystem code can be seen as a stabiliser code with an arbitrary partition of the logical qubits into both logical qubits and gauge qubits.
    The other definition of a subsystem code, via a gauge group, is illustrated with a grey box.
    (Note - this representation ignores phases, which we would need to include if we were to write a full stabiliser tableau---see \cite{nielsenQuantumComputationQuantum2010})}
\label{fig:symplectic}
\end{figure}

Assuming the dropouts are sparsely distributed across the device, we will still have a set of untouched stabiliser generators $U \subseteq Q$ that still commute with everything in $Q$.
The remaining `touched' quasi-stabilisers $T = Q \setminus U$ need to be packaged into a maximal set of product stabilisers $P$, specified by a binary matrix $B$, and a set of exclusively anticommuting gauge operators $G$ specified by a binary matrix $C$ with 
\begin{align}
   p_i = \prod{t_j^{B_{ij}}} &&\text{and} && g_i = \prod{t_j^{C_{ij}}}.
\end{align}
Finally, we need a set of logical operators $L$ that commute with all the operators defined so far, so that (subdividing $G = G^X \sqcup G^Z$ and $L = L^X \sqcup L^Z$) the following commutation rules hold for all $i$ and $j$:
\begin{align}
    [u_i,u_j] = [u_i,p_j]&=  0 \\
    [g^X_i,g^Z_j] = [l^X_i,l^Z_j] &= \delta_{ij} \\
    [u_i,p_j] = [u_i,g_j] = [p_i,g_j] = [u_i,l_j] &= [p_i,l_j] = [g_i,l_j] = 0 
\end{align}
where I have abused notation slightly in defining $[a,b] = 0$ where $a$ and $b$ commute and $1$ where they anticommute.
These admittedly dense relations are represented visually in Figure \ref{fig:symplectic}, where I also split our operator sets $U$ and $P$ into $X$ and $Z$ parts.

To construct all these operators, we first compute the binary anticommutation matrix $A$. $A$ has $a^X$ rows, equal to the number of $X$-type operators in $T$, and $a^Z$ columns, equal to the number of $Z$-type operators in $T$.
We set $A_{ij} = 1$ iff $t_i$ anticommutes with $t_j$, so that we can view $A$ as the biadjacency matrix of an `anticommutation graph' as shown in Figure \ref{fig:overview}.
We may then represent products of the $X$-type and $Z$-type quasi-stabilisers as binary vectors $v \in F_2^{a^X}$ and $w \in F_2^{a^Z}$, so that $vAw^T = 0$ if the corresponding product operators commute and $vAw^T = 1$ if they anticommute.
We can bidiagonalise $A$ using Gaussian elimination to obtain
\begin{equation}
    VAW^T =
\begin{bmatrix}
I_g & 0 \\
0   & 0
\end{bmatrix}
\end{equation}
where $g$ is the rank of $A$, and $V$ and $W$ are invertible matrices.
Then the first $g$ rows of $V$ and of $W$ are the exclusively anticommuting gauge operators we're looking for, expressed as products of quasi-stabilisers. In other words, stacking the first $g$ rows of $V$ and $W$ gives us the matrix $C$.
The remaining rows of $V$ and $W$ are product stabilisers, forming the rows of the matrix $B$.
We can then multiply out the definitions of the quasi-stabilisers to obtain the parity check matrices $H^X$ and $H^Z$ and gauge operator matrices $G^X$ and $G^Z$.
This set of product stabilisers is maximal and its span is unique.
Although in principle error correction performance could be improved by choosing `nicer' bases for the product operators, the algorithm is observed to generally produce product stabilisers that have low weight.

Once we have our stabilisers and gauge operators, we can construct the $X$ (resp. $Z$) logical operators of the code by finding row vectors that commute with $H^X$ and $G^X$ (resp. $H^Z$ and $G^Z$) operators whilst excluding the span of $H^Z$ (resp. $H^X$).
This can be done with a few applications of Gaussian elimination.

To measure a product stabiliser $p_i = t_a t_b ...$, we may measure the constituent quasi-stabilisers simultaneously or sequentially.
Either way, we require that at some point we collapse the qubits into a simultaneous eigenstate of the constituent operators. 
This is straightforward if the constituents are measured simultaneously (i.e. during the same contraction layer), but if they are measured sequentially, then we must ensure no quasi-stabiliser that anti-commutes with the constituents is measured in between the measurement of the first and last operators, lest the qubits be taken out of the shared eigenspace of the already-measured quasi-stabilisers.
I show how to encode these requirements as an integer linear programming problem in Section \ref{sec:sat}.

\subsection{Producing contraction schedules and finding scheduling constraints}

The ancilla-free framework for error-correcting codes allows for a great deal of flexibility in how the mid-cycle stabilisers are measured.
In this section, I describe a routine that takes as input a global connectivity graph and a set of quasi-stabilisers (Section \ref{sec:subsystem}), as well as a desired contraction step length $t$ (e.g. $t=2$ layers of CNOTs for the surface code) and produces an abstract scheduling graph.
The nodes of the abstract scheduling graph represent quasi-stabilisers, each of which is decorated with a list of possible schedules for contracting that node, and each of whose edges is decorated with pairs of compatible schedules between the quasi-stabilisers represented by its two endpoints.
The intra-layer constraints encoded in the scheduling graph, combined with the inter-layer requirements resulting from the structure of the subsystem code (Section \ref{sec:subsystem}), give us the full SAT problem (Section \ref{sec:sat}) we solve and optimise over to give us full multi-layer schedules that define a syndrome extraction round.

The scheduling problem is highly symmetric: all stabiliser codes considered in this work consist of a limited number of stabiliser `shapes' that tile to produce a code.
In my implementation, I leverage this fact by caching these shapes and calculating both the contraction schedules  and the constraints between them only for unique cases.
Here, I simplify the presentation, and provide these details \href{https://github.com/riverlane/ACID}{via the source code}.

First, we wish to calculate every possible contraction schedule for each quasi-stabiliser.
Each quasi-stabiliser has an associated `local connectivity graph', which is the subgraph of the global connectivity graph induced by its nodes.
A contraction schedule is a series of CNOTs, each on some edge of the local graph, over $t$ timesteps, that contracts a Pauli so that it is supported on only one root qubit.

Contraction schedules can be uniquely specified by 1) a rooted-spanning-tree subgraph of the local graph, and 2) a `timing': for each edge $e$, a timestep $1 \le t_e \le t$, such that no two edges that share a node are contracted at the same time, and contractions over edges more distant from the root occur before those closer to the root.
In the case of $X$-type quasi-stabilisers, this means that the first CNOTs to occur are targeted on leaf nodes of the tree, and remove the support of the Pauli string on the leaf-node qubits.
Subsequent CNOTs then recede the support further up the tree until it is localised at the root.
Note that some rooted spanning trees have no valid timings.
The spanning trees can be enumerated in linear time using a standard algorithm \cite{winterAlgorithmEnumerationSpanning1986}.
The timings can also be enumerated in linear time using a simple recursive algorithm (see the source).

We then consider every pair of quasi-stabilisers $q_a, q_b$ that share at least one qubit.
For each pair, we consider every pair of contraction schedules $c_a,c_b$, and decide if they are compatible according to the following criteria:
\begin{enumerate}
    \item the CNOTs in $c_a$ and $c_b$ act on disjoint qubits in each timestep, excepting the case where the two CNOTs are the same operation.
    \item Neither schedule is permitted to modify the support of Pauli string of the other quasi-stabiliser. For example, an $X$-type quasi-stabiliser $q_a$ sharing a single qubit with another $X$-type quasi-stabiliser $q_b$ could not have a schedule $a$ containing in the first timestep a CNOT controlled on the common qubit, as this would copy the support of the $q_b$'s Pauli string into itself, preventing $b$ from being able to contract the string onto the chosen root qubit.
\end{enumerate}

To enforce the second criterion, we precalculate the evolution of the Pauli string through each schedule and compare it with the CNOTs of other schedules as necessary.
Any pair $c_a,c_b$ satisfying the above will contract the Pauli strings onto their respective roots simultaneously. 
Note that any two anticommuting quasi-stabilisers can never have any compatible contraction schedules, as the contraction process is unitary and hence preserves (anti)commutation of observables; but the contraction process is required to `separate' the Pauli strings onto disjoint root qubits, which always commute.
The scheduling graph will in these cases have an edge connecting the two quasi-stabilisers decorated with an empty list containing no allowed schedules.

By checking all pairs against these criteria, ACID produces a `scheduling graph' that represents all the constraints that prevent two quasi-stabilisers from being contracted simultaneously with given schedules.

\subsection{Constructing an integer linear programming problem} \label{sec:sat}

Having constructed a subsystem code from our defective stabiliser code and calculated all the physical constraints on contraction schedules for neighbouring quasi-stabilisers, ACID builds an integer linear programming (ILP) problem that can be fed into a solver to find schedules.

Ideally, the final schedule will use as few layers per round as possible, so we start by constructing a problem with $L=2$ layers, and increase $L$ if the problem is found to be infeasible by the solver.

ACID defines variables corresponding to each schedule on each quasi-stabiliser on each layer, and a further set of variables that encode the product stabilisers of our subsystem code (Section \ref{sec:subsystem}).
It then encodes into our problem (see Appendix \ref{sec:sat-details} for details) the constraints:
\begin{itemize}
    \item no two schedules are chosen on neighbouring quasi-stabilisers within the same layer that are incompatible (physical constraints);
    \item no quasi-stabiliser is measured that would randomise the result of the measurement of a product stabiliser currently being measured.
\end{itemize}

ACID solves the problem using CP-SAT \cite{GoogleOrtools2025}, which is known to handle large-scale scheduling problems well.
The solver is given the task of trying to minimise the number of layers required to measure all the stabilisers at least once over a round.
As a secondary heuristic optimisation, it tries to maximise the number of times each stabiliser is measured.
The output of the solver gives us $L$, and a `global schedule'---a recipe for which quasi-stabilisers are to be measured in which layers and using which contraction schedule.

\subsection{Assigning detectors}\label{sec:detectors}

\begin{figure}[H]
    \centering
    \includegraphics[width=1\linewidth]{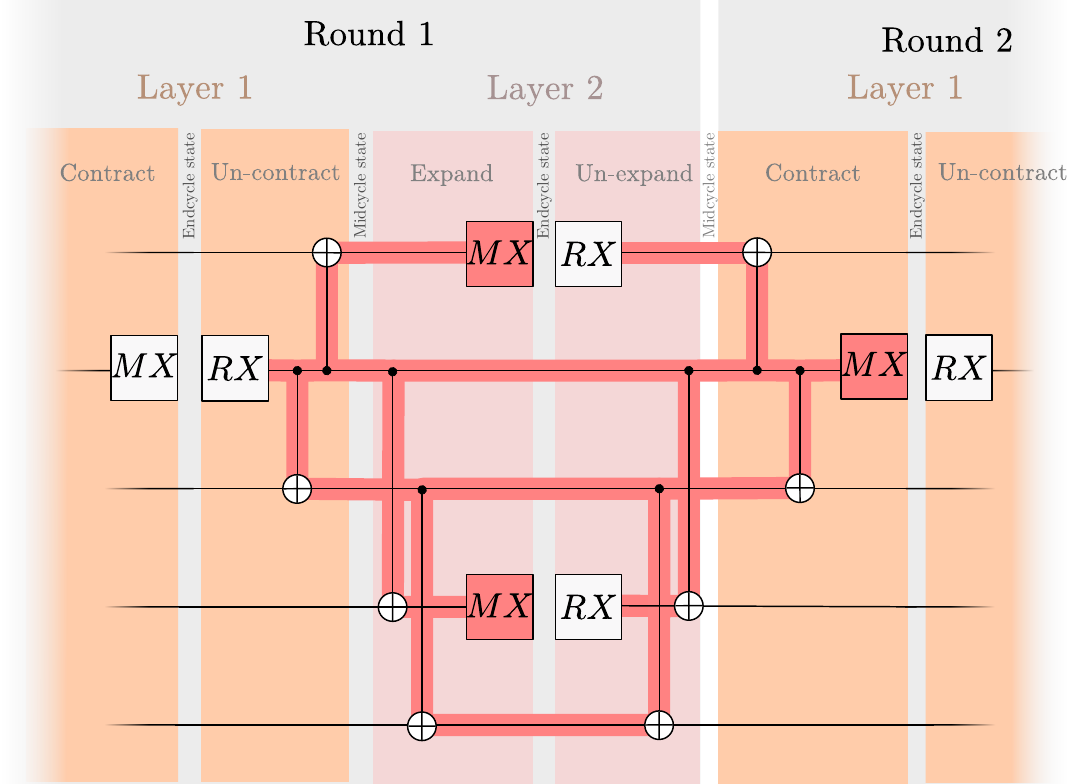}
    \caption{\textit{Detector assignment in the ancilla-free paradigm.}
    Depicted is a portion of two rounds of syndrome extraction, each of which has $L=2$ layers.
    A detector is defined between two rounds, being contracted in the first layer of both.
    The detector includes the final measurement, but also measurements of the stabilisers expanded support in the intervening second layer.}
    \label{fig:detector}
\end{figure}
The primary output of ACID is a syndrome extraction circuit, which in practice would be performed at regular intervals throughout a quantum computation.
To test its performance in the present work I simulate memory experiments consisting of $R$ rounds of syndrome extraction, where $R$ is set to the expected code-capacity distance of the end-cycle codes in the dropout-free case (see Section \ref{sec:bbcode} for the different codes).
Each of the $R$ rounds consists of $L$ layers wherein some subset of the quasi-stabilisers are contracted and measured.

Stim \cite{gidneyStimFastStabilizer2021b}, the Clifford simulator used here for the memory experiments, defines `detectors' as sets of measurements whose overall parity is deterministic when no errors occur: violations of these expected parities are reported to the decoder.
It is the task of the decoder to infer from the detector results which errors occurred in the circuit.

It is most helpful to think of these detectors in terms of Pauli flows \cite{browneGeneralizedFlowDeterminism2007}.
For our purposes, Pauli flows are Pauli strings that propagate through our circuit in the usual way, where we can add to the support of the Pauli string (`seed the flow') and where we can reduce the support of the Pauli strings (`terminate the flow') on measurements in the correct basis.
Any Pauli flow that is both seeded and terminated within the circuit can be used to define a valid detector.

Consider the case of a simple non-product stabiliser consisting of a single quasi-stabiliser.
Suppose the stabiliser is contracted in layer $A$ and then again in layer $B$.
This case is illustrated in Figure \ref{fig:detector}.
Layers $A$ and $B$ may lie in different rounds of syndrome extraction or they may both lie in the same round if the scheduler has managed to include extra redundant contractions of the stabiliser within one round. 
(In the case of Figure \ref{fig:detector}, `layer $A$' refers to layer 1 of round 1, and `layer $B$' refers to layer 1 of round 2.)
This means that in both layers $A$ and $B$, the stabiliser is contracted to some root qubit, which is measured and reset, and then un-contracted.
In general, the root qubits onto which the stabiliser is contracted may not be the same in $A$ and $B$.
We imagine a Pauli flow seeded on the root qubit reset instruction in layer $A$.
The string, which starts as a single $X$ or $Z$, un-contracts in the second half of layer $A$ to have exactly the support of the stabiliser in the mid-cycle code.
In any layers between $A$ and $B$, the string expands, where in general some of the qubits in the support of the expanded operator are measured and reset.
Note that these measurements will always be in the correct basis---if one wasn't, it would imply that our stabilisers don't all commute.
Since the Pauli flow cannot terminate on these resets that occur during the expansion layers, we must include the measurements that precede them and re-seed the Pauli flow on the resets.
This allows the Pauli string to `flow through' the measurement and reset and then un-expand and contract onto a single qubit in $B$, where we terminate the Pauli flow by measuring that qubit.
ACID adds detectors of this form between each contraction and the next.

Defining detectors for product stabilisers, each made up of multiple anti-commuting quasi-stabilisers, is more complicated.
A single measurement of the value of a product stabiliser might take multiple layers.
Define a `completion' for a product stabiliser as a series of layers over which all of the constituent quasi-stabilisers of a product stabiliser are contracted and measured.
We are guaranteed by the solver to have at least one completion per round of syndrome extraction.
Detectors are then defined between each completion event and the next.
To define a detector between completion $A$ and completion $B$, we imagine a Pauli flow seeded on each root-qubit reset in the contractions that comprise completion $A$.
We track the Pauli string resulting from these seeds, noting that in general they interfere with one another.
Then as in the non-product case, we must terminate and re-seed the Pauli flow wherever the expanded support of the product stabiliser is measured.
Finally, we terminate the flow with the measurements of root qubits that correspond to the measurements comprising completion $B$.

We can also add some extra detectors where a quasi-stabiliser is measured more than twice over two completions, and there is no anticommuting quasi-stabiliser that randomises its value in between two of them.
Also note that to simulate a memory experiment, we add extra detectors between each contraction and the initial and final Pauli product measurements we use to put the qubits into the codespace and measure out at the end.

Stim observables, the circuit-level analogue of logical operators, are also defined as sets of measurements whose overall parity is deterministic in the absence of errors, and can be specified using Pauli flows.
In the language of ancilla-free QEC, we can think of a logical operator as an operator that is never contracted.
It is therefore expanded in every layer of every round, and we must track its expansion and terminate and re-seed the flow accordingly.
In general, expanding a stabiliser or observable can reduce its support, so contraction circuits can be distance-reducing.
For more details of the memory experiments I constructed for our simulations see Appendix \ref{sec:memory-experiment}.

\section {The surface code}

\begin{figure}
    \centering
    \includegraphics[width=1\linewidth]{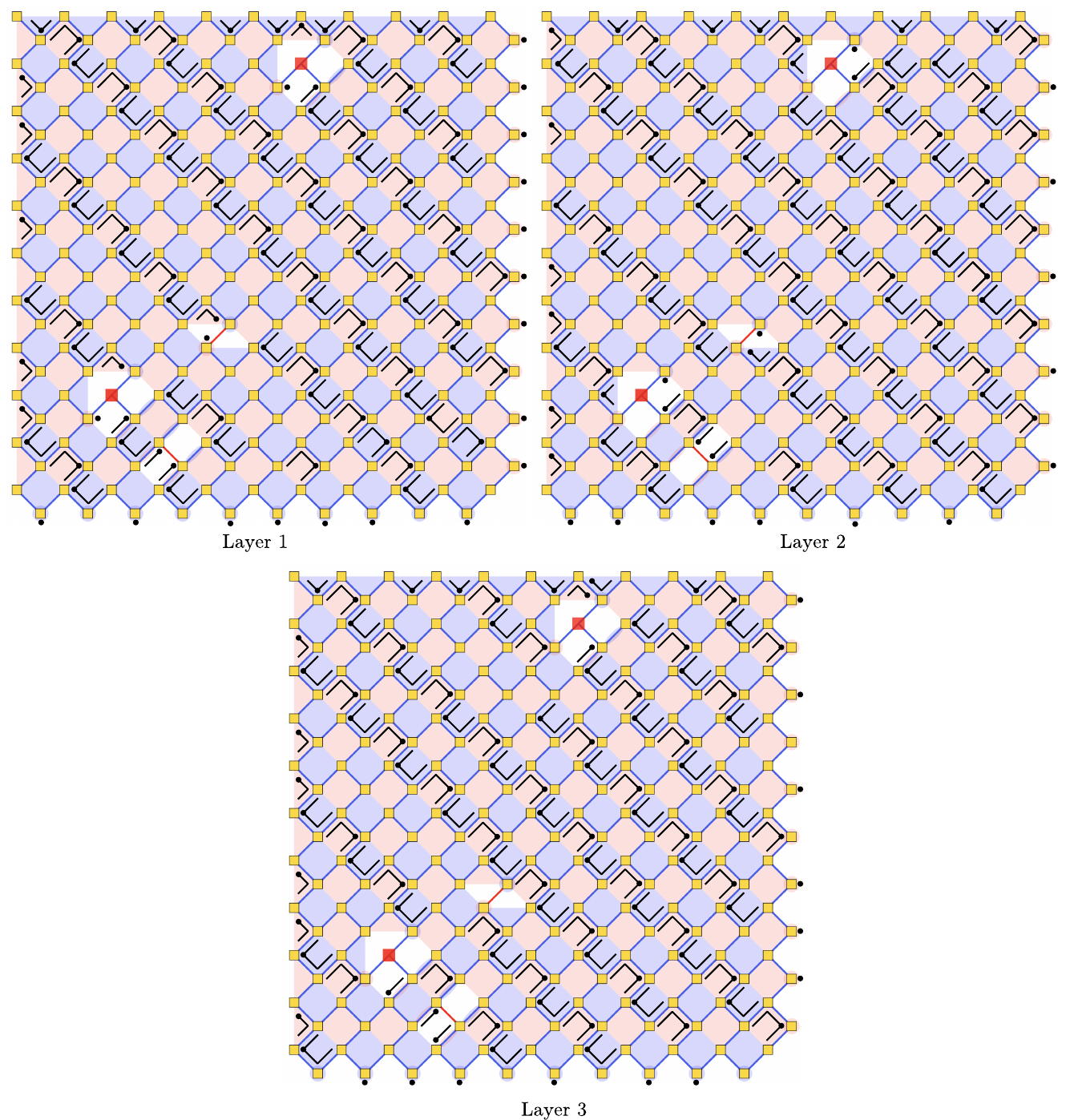}
    \caption{\textit{A LUCI-like diagram showing a syndrome extraction circuit of the brickwork surface code in the presence of two dropped qubits and two dropped couplers consisting of three contraction layers.} I've followed the notation of \cite{debroyLUCISurfaceCode2024}, where stabilisers that are contacted in a given layer are indicated with a black depiction of the rooted spanning tree used to contract them.
    Dropped qubits and couplers are indicated in red.
        \href{https://stasiu51.github.io/Shatter-Forward/}{Click here to open the circuit in Shatter.}
    }
    
    \label{fig:surface-example}
\end{figure}

The state of the art for handling dropouts in the surface code is the LUCI framework \cite{debroyLUCISurfaceCode2024}, which was the original inspiration for this work.
LUCI offers an intuitive way of measuring all quasi-stabilisers over $L=4$ contraction layers within a round of syndrome extraction.
The technique presented in the original paper relied on the flexibility in contraction schedules offered by a (degree-4) square-grid connectivity to deal with dropped couplers and resorted to forming product stabilisers to deal with dropped qubits, with certain pathological patterns of dropped couplers treated as dropped qubits.
The more recent work \cite{higgottHandlingFabricationDefects2025} allows LUCI to be applied to a degree-3 hexagonal/brickwork-style layout by assigning product stabilisers as soon as couplers are dropped (as well as some scheduling improvements).
In both works, couplers can often be dropped without reducing the code-capacity distance $d$ of the code.
In particular, in the square-grid case the loss of an isolated coupler can always be accommodated without losing distance, and in the hex-grid case we can tolerate the loss of about a third of the couplers without losing distance, with the distance reducing by one for the other couplers.

\begin{figure}
    \centering
    \includegraphics[width=0.78 \linewidth]{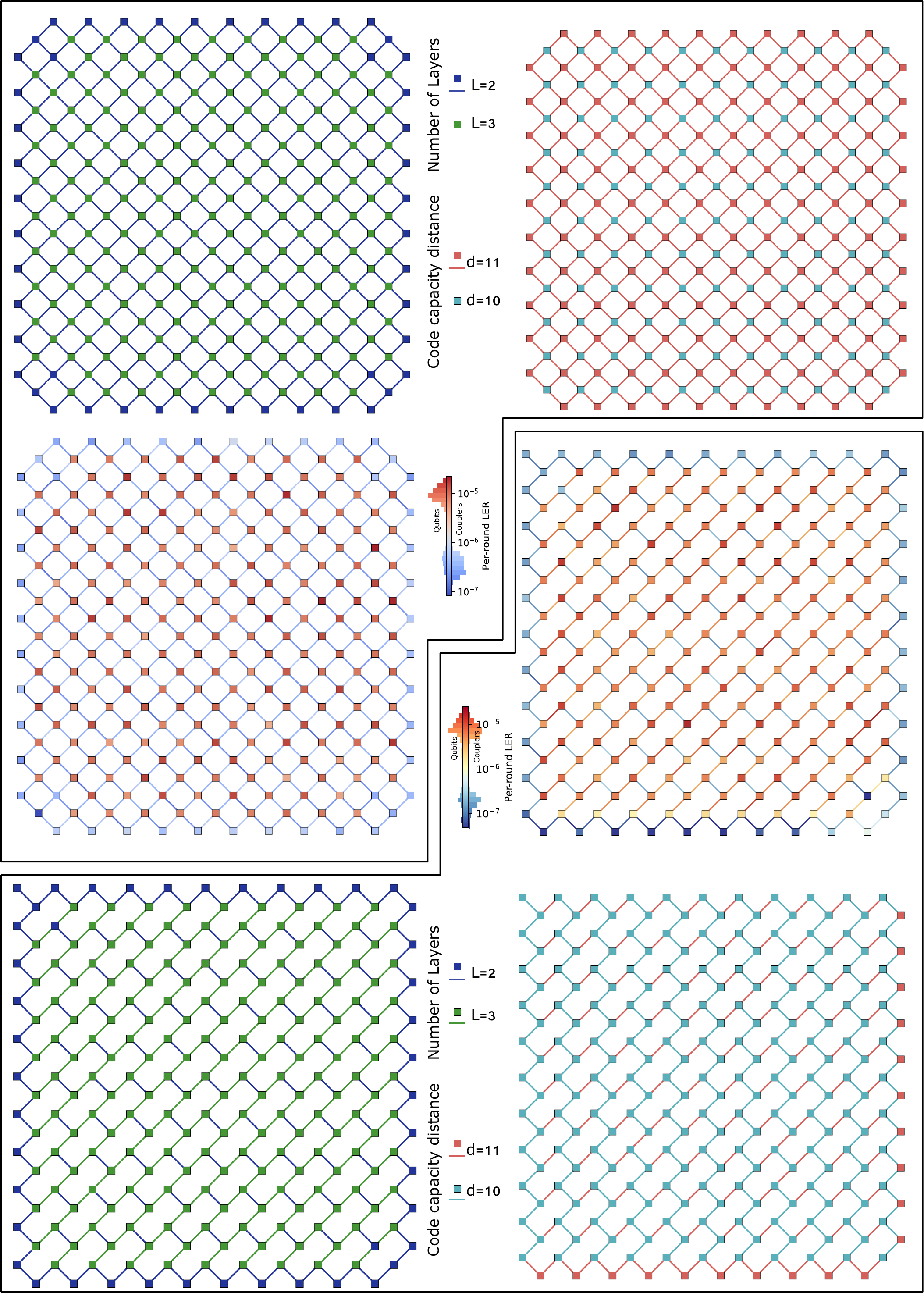}
    \caption{\textit{The effect of dropping a single qubit or coupler from the surface code in square-grid and hex-grid configuration.}
    The top-left half of the figure shows the effect on the number of layers required for a syndrome extraction circuit $L$ (top-left), the mid-cycle code capacity distance (top-right), and logical error rate (middle-left).
    The bottom-right half of the figure shows the same for the hex-grid surface code.
    In each case, each qubit and coupler is coloured according to the effect that the loss of that qubit or coupler has on the relevant quantity.
    For the logical error rate (LER), the central colour-bars are equipped with histograms on the left and right corresponding to the the distribution of the dropped-qubit (left) and dropped-right (right) LER distribution.
    }
    \label{fig:surface-viz}
\end{figure}

Applying ACID to both the square-grid and hex-grid layouts, I find that ACID can often compile a syndrome extraction circuit with $L=2$ layers and in all cases I tried (up to the loss of 3 qubits and 3 couplers), can create a circuit with $L=3$.
In principle this should allow the time between lattice surgery operations to be reduced from $4d$ using LUCI to $3d$ or even $2d$ with sufficient quality fabrication and device postselection.
In Figure \ref{fig:surface-viz}, we show the effect on $L$, $d$ and the logical error rate of dropping out any single coupler or qubit in a distance-11 patch of square-grid and hex-grid surface code.
As with LUCI, the code-capacity distance (Appendix \ref{sec:distance}) is often preserved, with the square-grid connectivity unsurprisingly more tolerant to dropped couplers than the hex-grid connectivity.
The relative performance of LUCI and ACID is summarised in Figure \ref{fig:surfacesummaryplot}, and
I give a full characterisation of the performance of ACID and LUCI under different numbers of dropped qubits and couplers in Appendix \ref{sec:plots}.

An important caveat is that ACID sometimes produces detector error models that do not admit ready approximation as decoding graphs, meaning we lose access to fast matching decoders.
I believe that this is a result of overlapping redundant measurements of quasi-stabilisers, and can be fixed in future work by imposing additional constraints on the scheduling graph.

\section{Bivariate bicycle codes}\label{sec:bbcode}

\begin{figure}[h]
    \centering
    \includegraphics[width=1\linewidth]{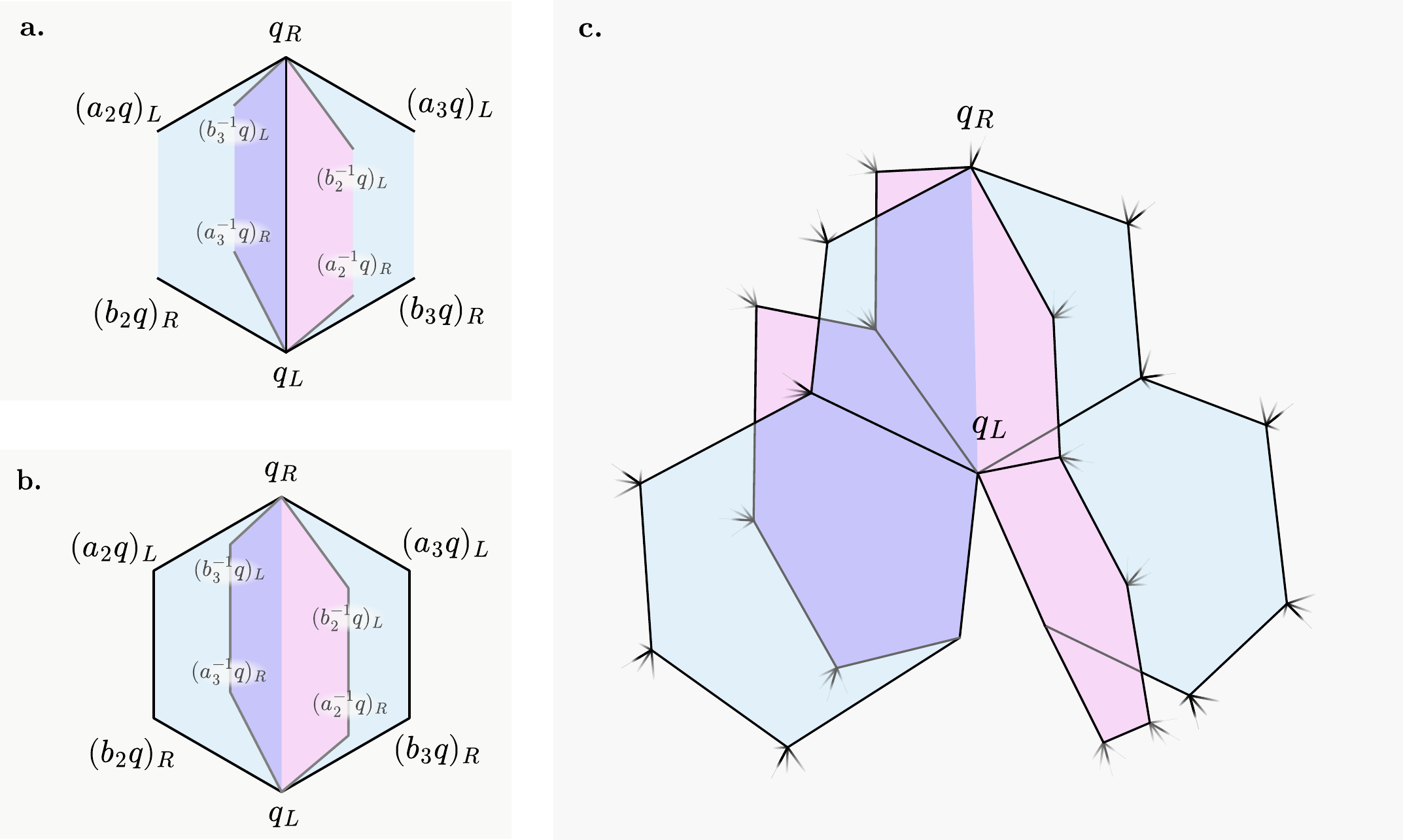}
    \caption{\textit{Connectivities for the bivariate bicycle (BB) code.}
    Vertices are qubits, and solid lines between them represent the presence of couplers.
    I've used the standard polynomial specification of BB codes where qubits are labelled as $q_{L/R}$, with $q$ a monomial in a finite polynomial ring and a left (`L') or `right' (`R') label.
    \subfiglab{a.} The local connectivities of $X$ (blue) and $Z$ (pink/purple) stabilisers in construction taken from \cite{shawLoweringConnectivityRequirements2025}.
    This construction results in 5-regular global connectivity graph.
    \subfiglab{b.} The local connectivity of the same stabilisers in my hexagonal connectivity. Note that the qubits associated with monomials $q_L$ and $q_R$ are no longer connected, but we introduce two new edges with directions $a_2 b_2^{-1}$ and $a_3 b_3^{-1}$.
    \subfiglab{c.} All of the stabilisers containing a given $L$ qubit in the hexagonal connectivity, showing the 6-regularity of the graph.
    }
    
    \label{fig:hexconn}
\end{figure}

Notwithstanding discussions of possible theoretical methods for doing so \cite{xuFaulttolerantProtocolsSpacetime2025, zhouLouvreRelaxingHardware2025}, I believe ACID is the first concrete proposal for handling dropouts in the bivariate bicycle (BB) codes \cite{bravyiHighthresholdLowoverheadFaulttolerant2024b}.
Following \cite{shawLoweringConnectivityRequirements2025}, I applied ACID to the 144-qubit and 288-qubit codes, each with 12 logical qubits, from \cite{bravyiHighthresholdLowoverheadFaulttolerant2024b} as mid-cycle codes.
In the dropout-free case, these mid-cycle codes produce end-cycle codes with distance 6 and distance 12 respectively (these are the distances I use as the length of the memory experiments).

I applied ACID to these codes with two underlying connectivities, inspired by the hex-grid and square-grid surface code connectivities: first, the degree-5 connectivity introduced in \cite{shawLoweringConnectivityRequirements2025}, and second a `hexagonal' connectivity in which the six qubits defining a stabiliser are connected in a cycle (see Figure \ref{fig:hexconn}).
This latter connectivity is degree-6, but offers more flexibility in contraction schedules for each stabiliser.
In particular, where in the degree-5 connectivity the loss of any coupler results in anticommuting quasi-stabilisers (as the local connectivity graph is a tree), the cyclic nature of the hexagonal connectivity means we can tolerate the loss of isolated couplers whilst still being able to measure all the original stabilisers.
It is interesting to note that for some BB codes, such as the 288 code, the hexagonal connectivity is completely symmetrical in its couplers and qubits (the associated undirected graph is arc-transitive), whereas in others such as the 144 code the edges split into two or more edge orbits.

The non-locality and complexity of these connectivities mean that neither will likely prove a viable layout for practical quantum hardware.
Future work should examine the dropout tolerance of proposals such as \cite{eberhardtPruningQLDPCCodes2024, liangPlanarQuantumLowDensity2025} that aim for more hardware-friendly QLDPC codes.

The full data for the 144-qubit and 288-qubit BB codes for up to 3 dropped qubits and couplers is shown in Appendix \ref{sec:plots}. 
For the hexagonal connectivity, in all the scenarios tested, the solver finds a schedule with $L=4$ or fewer.
Using the more limited degree-5 connectivity, the solver occasionally needs $L=5$ for the 144 code.

Of particular interest is the single-dropped-coupler case for the 144-qubit code with the hexagonal connectivity (Appendix \ref{sec:plots}).
I note that the introduction of a single defective coupler results in most cases in a substantially lower logical error rate than in the dropout-free case.
Investigating further, one can calculate the distance of the end-cycle codes resulting from a given schedule and see that certain contraction schedules lead to an end-cycle-distance of 7 or even 8.
Alternative contraction schedules with higher distances are also discussed (in the degree-5, dropout-free case) in the appendix of \cite{shawLoweringConnectivityRequirements2025}.
It is noteworthy that the dropout-free schedules are tightly bunched in logical error rate, suggesting that the heuristic metrics that the solver optimises for in fact push it into a schedule that is sub-optimal, and that the frustration introduced by a new constraint allows the solver to explore better options.

Whereas for the surface code lattice surgery is straightforwardly generalised to the case of patches with dropouts, applying ACID to the BB codes corrodes the symmetry of the code that is used in proposals for logical processing with the code.
In particular, the code is no longer translationally invariant, so we appear to lose access to a large class of automorphisms used in \cite{yoderTourGrossModular2025} to perform universal computation.

\section{The colour code}\label{sec:colour}

I applied ACID to three flavours of colour code, with all data shown in Appendix \ref{sec:plots}.
The first, the degree-3 ancilla-free colour code is taken directly from \cite{gidneyNewCircuitsOpen2023a} (where it is called the `middle-out' circuit).
This connectivity is naturally implemented on a hex-grid qubit layout.
The local connectivity graphs in this connectivity are then hexagons, so we expect to often be able to measure the original stabilisers in the presence of dropped couplers.
However, the tightly packed and overlapping nature of the colour code stabilisers means that the global scheduling problem is very sensitive to local restrictions, and a single dropped coupler is normally enough to force the scheduler to require four contraction layers.

The second connectivity I tried is what results if you take the same middle-out colour code circuit but lay out the qubits on a full square lattice, meaning we introduce an extra redundant edge crossing from the two extremes of each local hexagon.
This `square-grid ancilla-free' colour code is the example used in Figure \ref{fig:overview}.
The extra redundancy introduced by this edge allows the solver to tolerate a single dropped coupler in three or fewer layers in the majority of cases.

Finally, I tried a version of the colour code based on the `superdense' syndrome extraction circuit also from \cite{gidneyNewCircuitsOpen2023a}.
Eight qubits are assigned to each stabiliser of the colour code, six outer qubits and two inner ones (these inner qubits were previously measurement ancillae).
I then give ACID the task of scheduling four stabilisers per stabiliser of the colour code: an $X$ and a $Z$ weight-8 stabiliser and $X$ and $Z$ weight-2 stabilisers supported on the inner qubits. 
The natural solution found by the solver is then almost exactly the superdense syndrome extraction circuit from the literature, but whereas the original superdense circuit has three `incoming' CNOTs preceding three `outgoing' CNOTs, in this circuit the two groups take turns to go first.
This leads to the same number of detectors, but they are more localised in time, suggesting the decoding performance may be better.
This superdense colour code is embedded on a full square-grid lattice.
This means we have two redundant edges in the local connectivity graphs of the weight-8 stabilisers.
This redundancy, combined with the fact that the contraction circuits take four timesteps, can lead to a combinatorial explosion in the number of contraction schedules, overwhelming the solver (even though in principle having lots of schedules to choose from is a good thing). 
To overcome this, we prune the contraction schedules ahead of feeding them to the solver based on their compatibility with certain `preferred schedules'.

\section{Outlook}

ACID can be applied to any CSS stabiliser code with any connectivity.
Even once components have dropped out, there are many possible syndrome extraction circuits to choose from.
ACID uses crude heuristics as tie-breaks, preferring global schedules that `cram in' as many stabiliser contractions as possible.
Some global schedules will needlessly reduce the circuit distance of the quantum memory circuit when others compatible with the same dropped qubits and couplers don't.
This is best illustrated in the case of the 144-qubit bivariate bicycle code, where the introduction of a single dropped coupler forces the solver to output global schedules with a lower logical error rate than those it finds in the dropout-free case.
Future work should address this, introducing additional constraints, analytic or heuristic, into the solver to force it to select better schedules.
Other requirements could be introduced in this way: in principle, one could vary the syndrome extraction circuit between rounds, measuring and resetting all qubits with some frequency in order to mitigate leakage.

Whilst the basic `merge and split' operations of surface code and colour code lattice surgery \cite{litinskiGameSurfaceCodes2019f, landahlQuantumComputingColorcode2014} should still be valid in the presence of dropouts, it is not clear how the loss of symmetry of codes in the presence of dropouts will affect logical operations for more general codes.

This work also has implications for the co-design of quantum hardware and error-correcting codes.
Different hardware connectivities should be considered not just for the codes they support, but the flexibility they provide in case of imperfect fabrication.
There may be design elements that can be introduced that substantially improve this flexibility without a correspondingly large overhead in qubit count or connectivity.

\section{Acknowledgements}

I'm hugely grateful to my supervisors, Dan Browne and Earl Campbell, for their support and sage advice throughout the completion of this work, and their comments on various drafts of this manuscript.

This work was supported by Riverlane and the Engineering and Physical Sciences Research Council [grant number EP/Y035046/1].

\printbibliography
\begin{appendix}

\section{Distance definitions}\label{sec:distance}

There are several definitions of distance relevant to our problem.
The simplest is the code-capacity distance of the mid-cycle code.
In the absence of dropouts, this is the weight of the lowest-weight logical operator of the code.
When dropouts are introduced, we construct a subsystem code (Section \ref{sec:subsystem}).
The relevant code-capacity distance is now the dressed distance, that is, the minimum weight of any operator
\begin{equation}
    L\cdot G
\end{equation}
where $L$ acts only on the logical qubits, and $G$ acts only on the gauge qubits, and we require that $L$ is non-trivial.
We can calculate the `dressed distance' of the subsystem code with $g$ gauge qubits by gauge-fixing.
For each of the $2^g$ binary vectors $z$ of length $g$, we construct a stabiliser code with an $X$ parity check matrix given by the parity check matrix of the code $H^X$ with the rows of $G^X$ corresponding to the 1s in $z$ appended, and a $Z$ parity check matrix constructed by appending the rows of $G^Z$ where $z$ is 0 to $H^Z$.
We find the distance of each of these codes using the QDistRand package \cite{pryadkoQDistRndGAPPackage2022}, and take the minimum.

We can also consider the distance of the end-cycle codes.
In the dropout-free case, this is useful as the distance of the end-cycle codes gives an upper bound on the overall circuit distance.
It is this code capacity distance of the end-cycle codes that we use to give the number of rounds $R$.
However, it is not clear that this is a useful concept once we're dealing with subsystem codes: in general the end-cycle code is also a subsystem code, but the stabilisers of the end-cycle code may derive from gauge operators of the mid-cycle code, if we have an anticommuting quasi-stabiliser in the mid-cycle code all of whose anticommuting neighbours are being contracted in a given layer.

We can define the circuit-level distance in the usual way as the minimum number of errors that must occur in the circuit to flip the value of an observable whilst not triggering any detectors.
This is very costly to estimate reliably, so in this work we focus on logical error rate instead, leaving a detailed examination of distance-reducing contraction layers to future work.

\section{Details of the CP-SAT problem}\label{sec:sat-details}

To enforce the quasi-stabiliser ordering constraints derived from the definition of product stabilisers (Section \ref{sec:subsystem}), we introduce FLAG, RESET, and COUNTER variables, each of which exists per-layer and has inter- and intra-layer constraints to model the scheduling we need.
\begin{itemize}
    \item The FLAG variables, one for each constituent quasi-stabiliser, track whether each constituent quasi-stabiliser has been `measured yet'.
    \item In a given layer, if all the constituent FLAGs are true, we set the boolean RESET flag variable on the next layer to true. If RESET is true on a layer, we set FLAG variables back to false (unless they are immediately re-measured), and increment a COUNTER integer variable associated with the product.
    \item Then we require that all $Z$-type product stabilisers $p^Z_a$ only start measuring once any $X$-type product stabilisers $p^X_b$ that contain any quasi-stabilisers with which any of $p^Z_a$'s quasi-stabilisers anticommute wait until the COUNTER variables associated with the conflicting $X$-type stabilisers are all greater than their own COUNTER variables.
    \item Finally, we enforce that any `orphan' quasi-stabilisers that do not form part of a product are also not measured at times they would interfere with the measurement of a product stabiliser.
\end{itemize}
Note that just because a quasi-stabiliser is not part of a product does not mean we can never form any detectors from it (Section \ref{sec:detectors}), but that these detectors are likely to be quite limited in coverage.

Having built this problem, we optimise as follows.
\begin{itemize}
    \item We require that all \textit{stabilisers} are measured at least once over the $L$ layers. Note that this includes both untouched stabilisers that did not suffer from defects rendering them anticommuting and product stabilisers formed from anticommuting quasi-stabilisers.
    \item We implicitly optimise over $L$ as a primary objective by constructing CP-SAT problems of increasing $L$, starting at 2, and reconstructing with higher $L$ if the solver deems the model INFEASIBLE. In our experiments, if $L$ is required to be higher than 5, we record a failure.
    \item We explicitly optimise, within each $L$, to \textit{maximise the  minimum number of times} any stabiliser is measured. This is based on the idea that if, say, we need $L=3$ layers to measure all the stabilisers but over those 3 layers we in fact measure each stabiliser twice, we may only need to wait $2d/3$ rounds between lattice surgery rounds to prevent harmful timelike errors. We leave examining this idea more closely to future work.
    \item Finally we tie-break between solutions by maximising the overall total number of quasi-stabilisers measured over all the layers.
\end{itemize}

The CP-SAT solver produces an assignment of a schedule (or none) for each quasi-stabiliser on each of the $L$ layers, which fully specifies a single round of a syndrome extraction circuit.

\section{Memory simulations} \label{sec:memory-experiment}

A quantum memory experiment should, as far as possible, simulate a portion of what would, in an actual quantum computation, be a period of qubit idling preceded and followed by either more idling or logical operations.
The Stim \cite{gidneyStimFastStabilizer2021b} memory experiments I ran to produce the plots in Appendix \ref{sec:plots} have the following stages.
For the noisy portion of the circuit, I assume $p=0.001$ probability of depolarisation after one- and two-qubit gates, and include bit- and phase-flips in the appropriate basis after resets and before measurements also with probability $p$.
\begin{itemize}
    \item First, we perform noiseless Stim Pauli-product measurements of all the quasi-stabilisers, to initialise the mid-cycle stabilisers of the code.
    \item Noiseless controlled$-L^X$ and controlled$-L^Z$ operations controlled on $2k$ noiseless reference qubits are performed to initialise $X$ and $Z$ observables.
    Concretely, for each $X$ logical, we perform CNOTs controlled on the reference qubit and targeting each qubit in the support of the logical operator.
    For the $Z$ logicals, the CNOTs are reversed.
    \item There are now $R$ rounds of noisy syndrome extraction.
    $R$ is set to the expected code-capacity distance of the end-cycle codes in the absence of dropouts (Appendix \ref{sec:distance}), and it is this $R$ that we use to calculate `per-round' logical error rate.
    Each syndrome extraction round contains $L$ contraction layers, each of which consists of $t$ timesteps of simultaneous CNOTs, measurement and reset of the root qubits, and then the same $t$ timesteps of CNOTs in reverse.
    \item The reference qubits are unentangled from the data qubits by noiselessly performing the same controlled-logical operations.
    \item Finally, we `close off' detectors by noiselessly measuring all the stabilisers with Stim Pauli product measurement instructions.
\end{itemize}

\section {Performance plots}\label{sec:plots}

All memory experiments were simulated at $p=0.001$ physical error rate and decoded with BP-AC \cite{wolanskiAmbiguityClusteringAccurate2025a}.

Each of the following plots is divided into panels.
In each panel I've fixed a number of qubits and couplers to drop out, and then sampled 30 dropout patterns, choosing uniformly among the qubits and couplers. 
ACID is then applied, producing a syndrome extraction circuit with $L$ layers.
In the background of each panel is a light-blue histogram showing how many of the dropout patterns resulted in a given $L$.
In the foreground, drawn on top of each histogram bar, is a red violin plot (essentially a smoothed histogram aligned vertically), showing the distribution of logical error rate per round (LER) for the dropout patterns, selected by $L$.
Syndrome extraction circuits consisting of more layers generally have a higher LER.

In the surface code plots, I include a comparison to a version of the LUCI algorithm, which always produces a syndrome extraction circuit with $L=4$ layers (Appendix \ref{sec:luci-repro}).

\section{Reproducing LUCI}\label{sec:luci-repro}

To compare ACID against LUCI \cite{debroyLUCISurfaceCode2024, higgottHandlingFabricationDefects2025} in the surface code, I implemented a version of the LUCI algorithm as a special case of ACID.
The product stabilisers, gauge and logical operators are found by the usual ACID routine, as are the local contraction schedules and the constraints between them.

Rather than tasking the solver with minimising the number of contraction layers required for a syndrome extraction circuit, we divide the surface code quasi-stabilisers into four `colours' 1-4 as described in the original LUCI paper, and require that the solver contract and measure the quasi-stabilisers of a given colour $i$ on layer $i$.
The solver is not required to use any particular schedule in measuring the required quasi-stabilisers, and is free to `fill in' additional quasi-stabiliser contractions on each layer as far as possible.
It is intended that this represents an upper bound on the flexibility of the LUCI algorithm.
This implementation does not explicitly attempt to construct `shells' as described in \cite{higgottHandlingFabricationDefects2025}.
However, it does attempt to measure quasi-stabilisers that form part of product stabilisers as much as possible, including `redundant' contractions that are not strictly required to measure product stabilisers, and it automatically assigns additional detectors in these cases, which likely amounts to the same thing.

\begin{figure}
    \centering
    \includegraphics[width=1\linewidth]{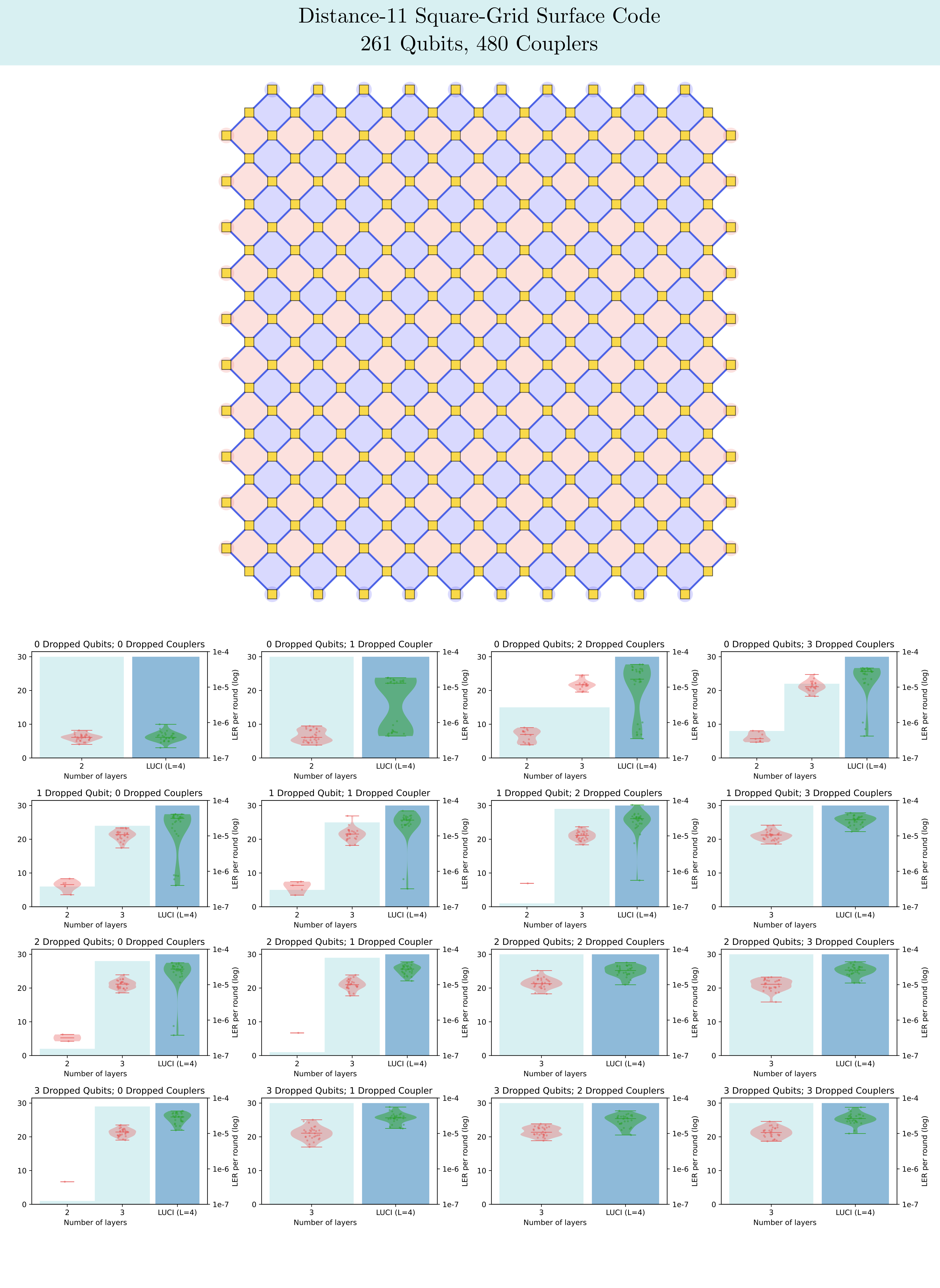}
\end{figure}
\begin{figure}
    \centering
    \includegraphics[width=1\linewidth]{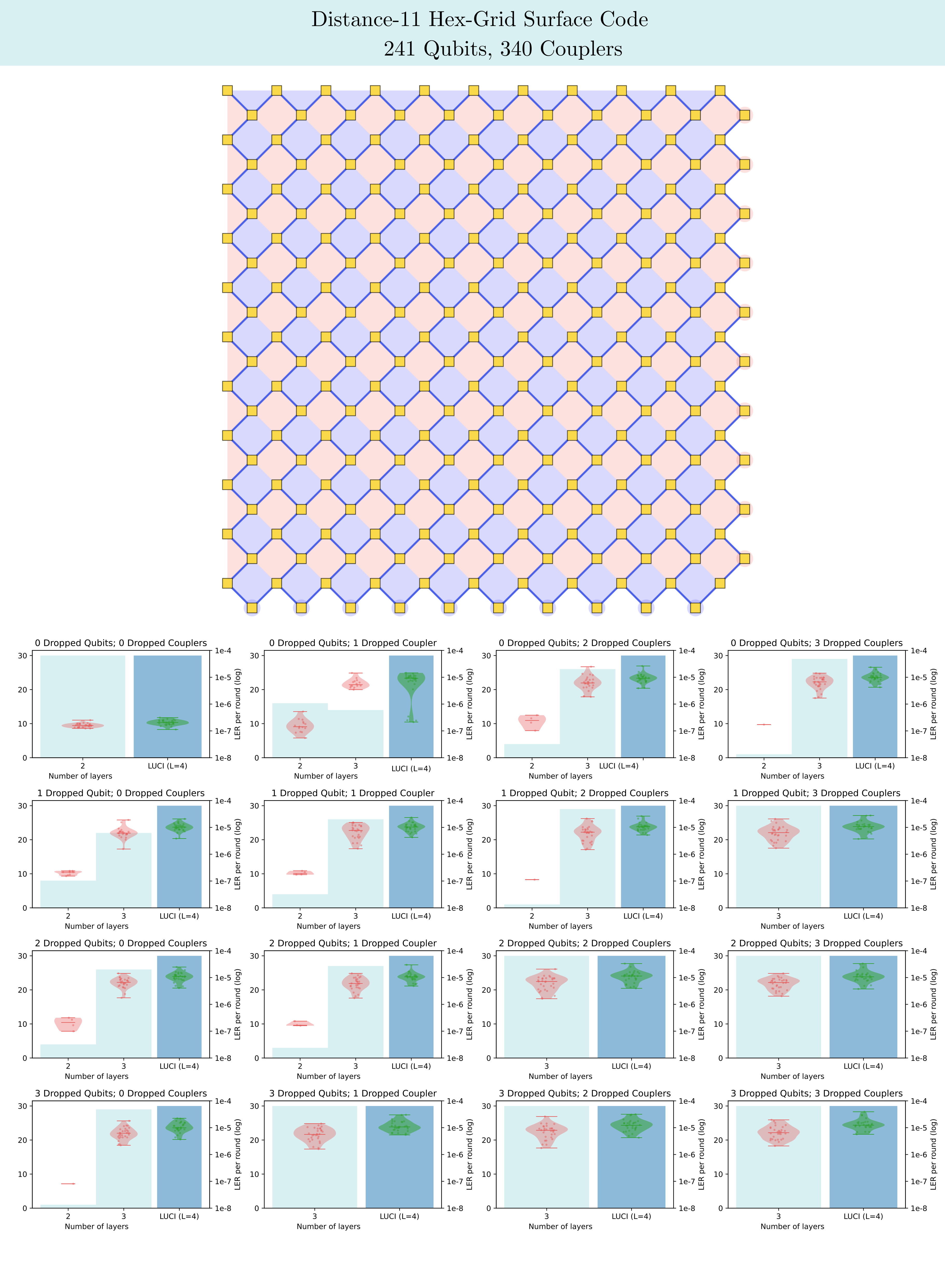}
\end{figure}
\begin{figure}
    \centering
    \includegraphics[width=1\linewidth]{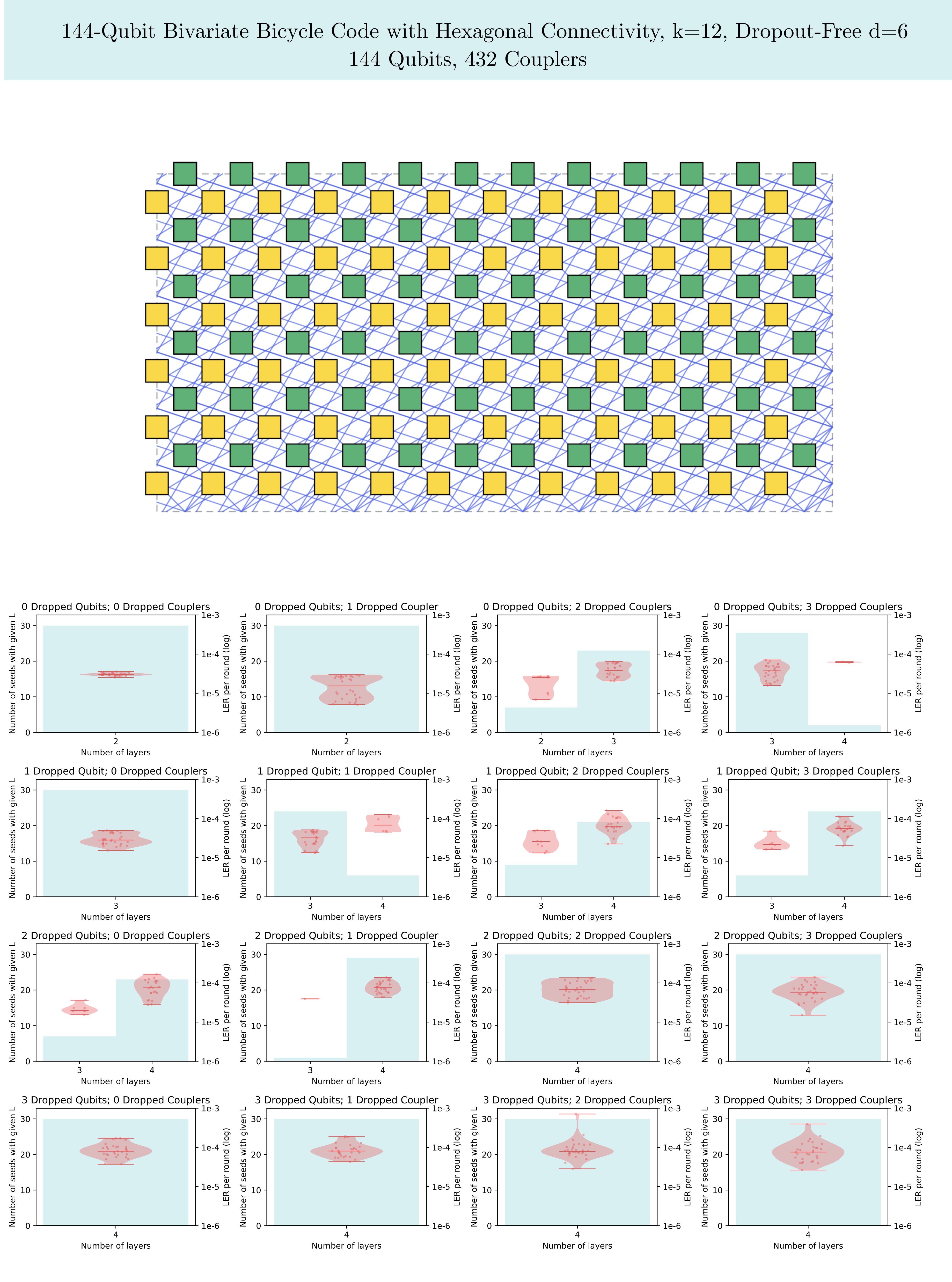}
\end{figure}
\begin{figure}
    \centering
    \includegraphics[width=1\linewidth]{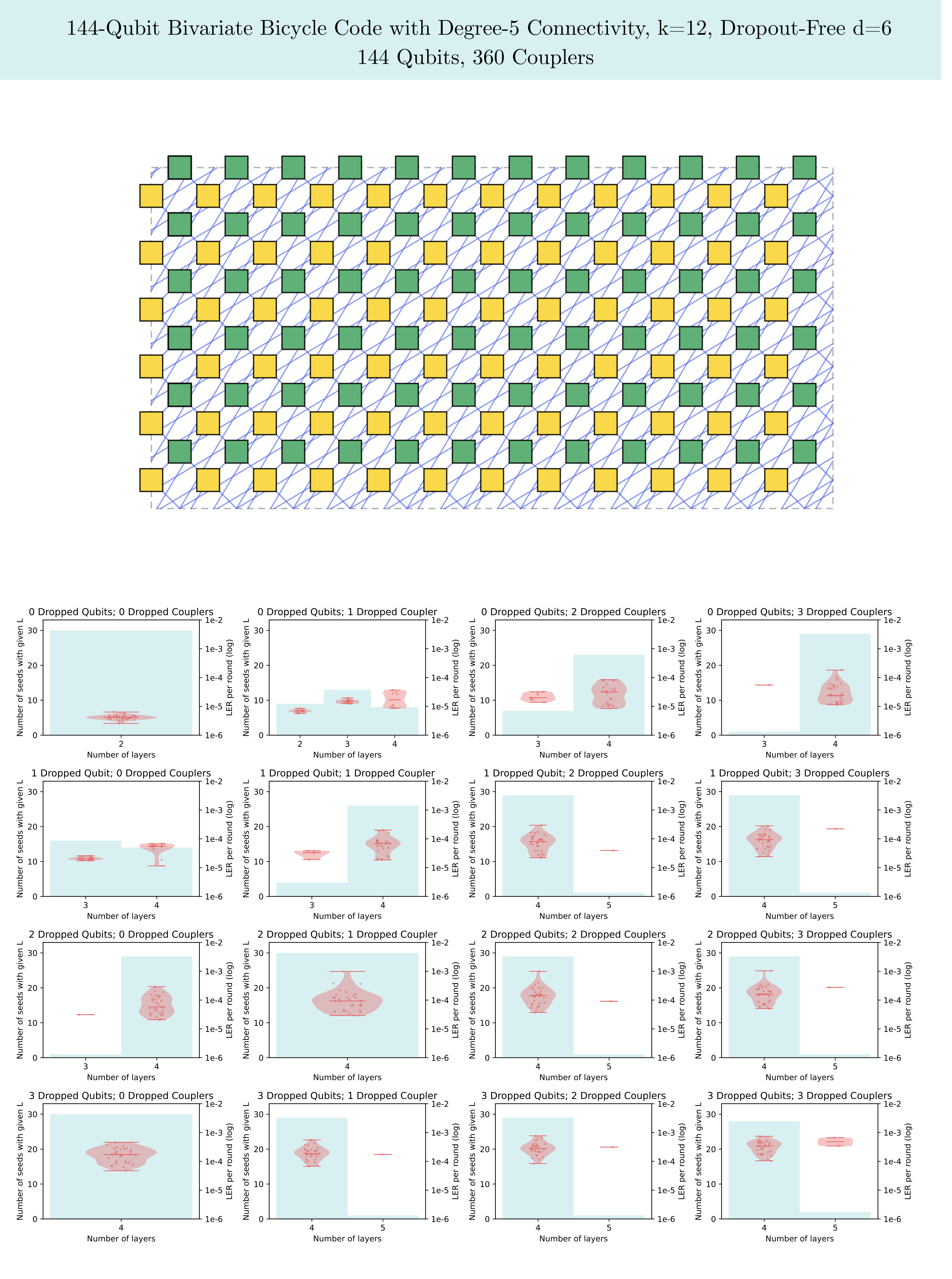}
\end{figure}
\begin{figure}
    \centering
    \includegraphics[width=1\linewidth]{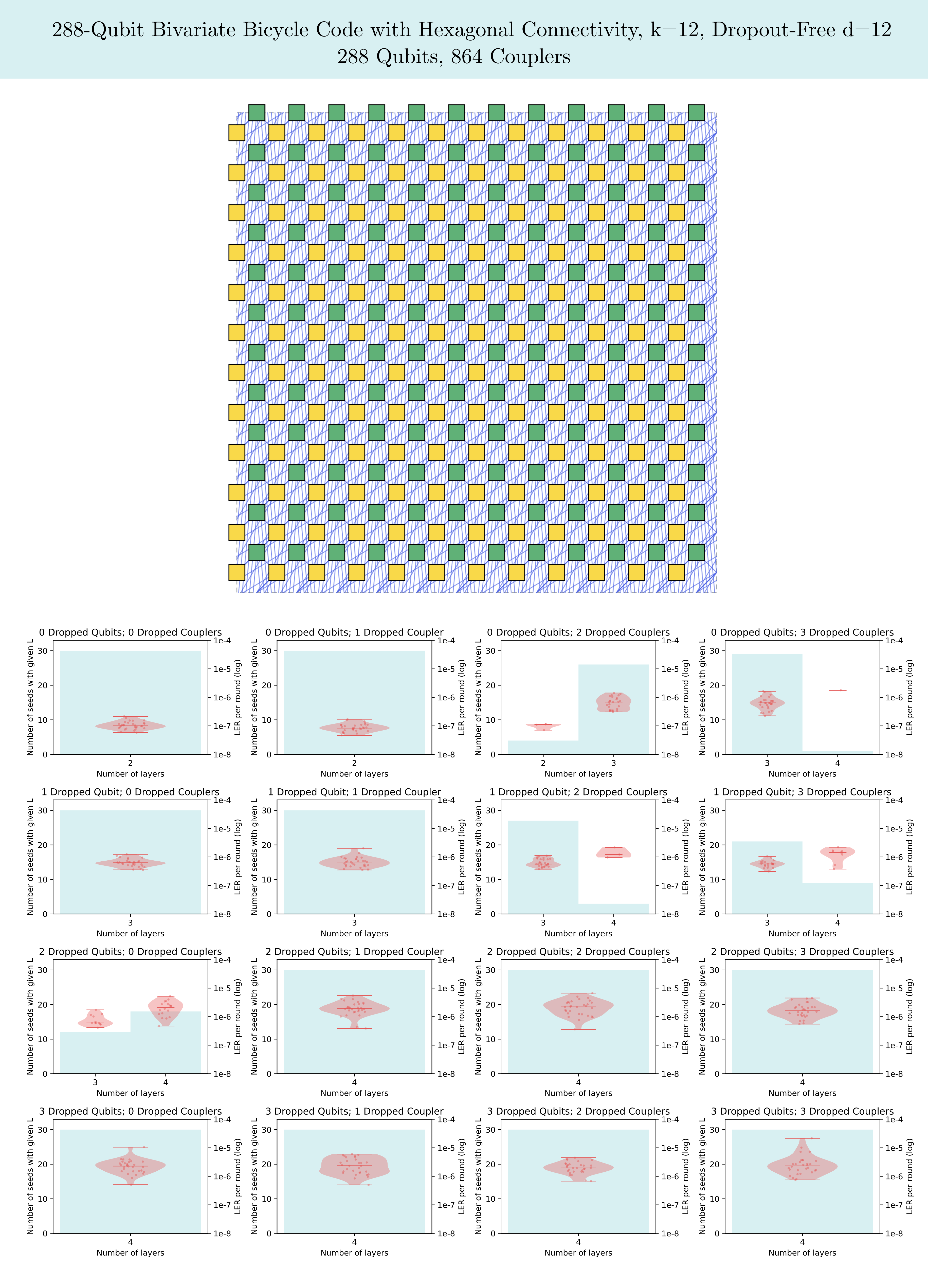}
\end{figure}
\begin{figure}
    \centering
    \includegraphics[width=1\linewidth]{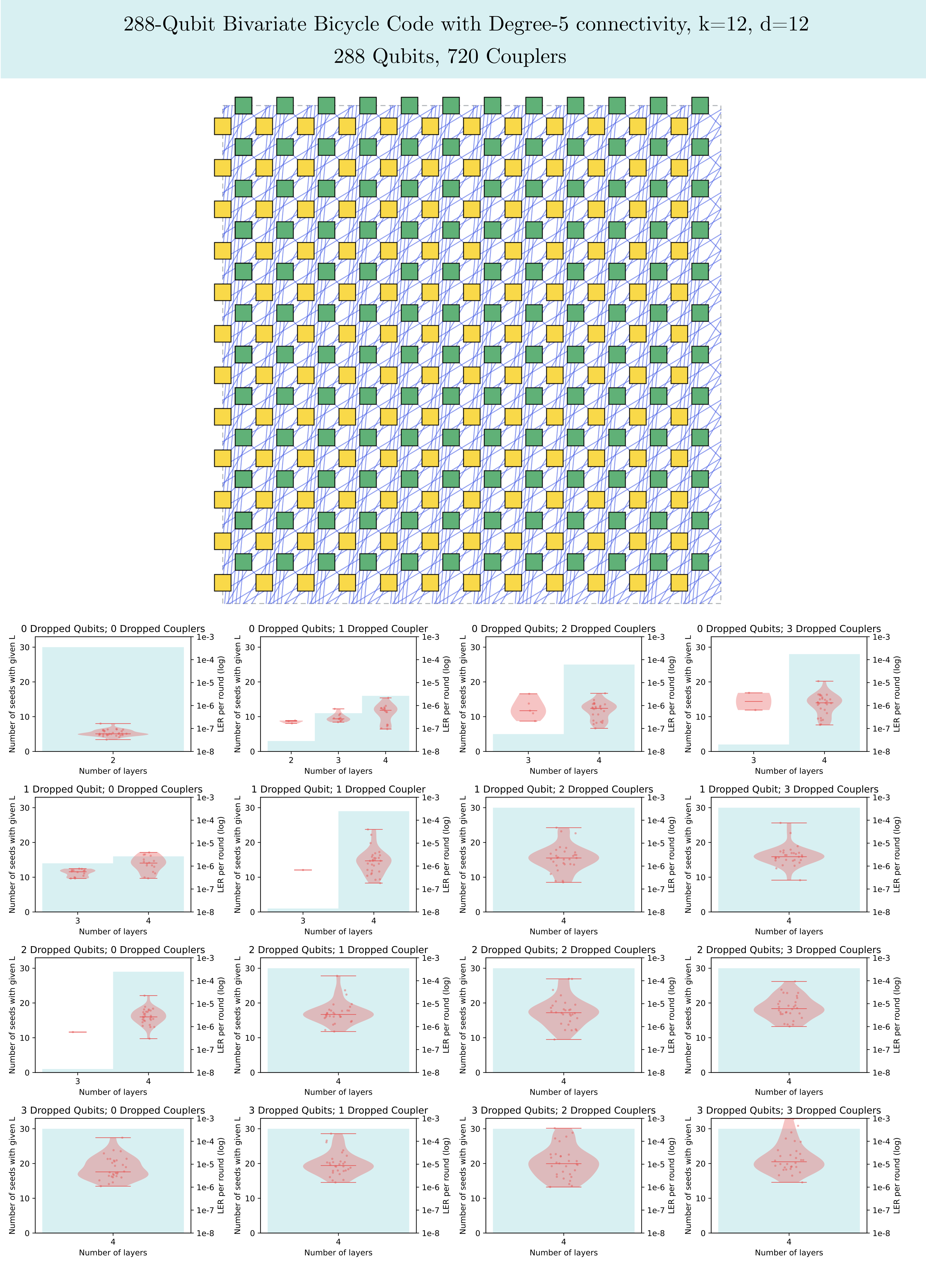}
\end{figure}

\begin{figure}
    \centering
    \includegraphics[width=1\linewidth]{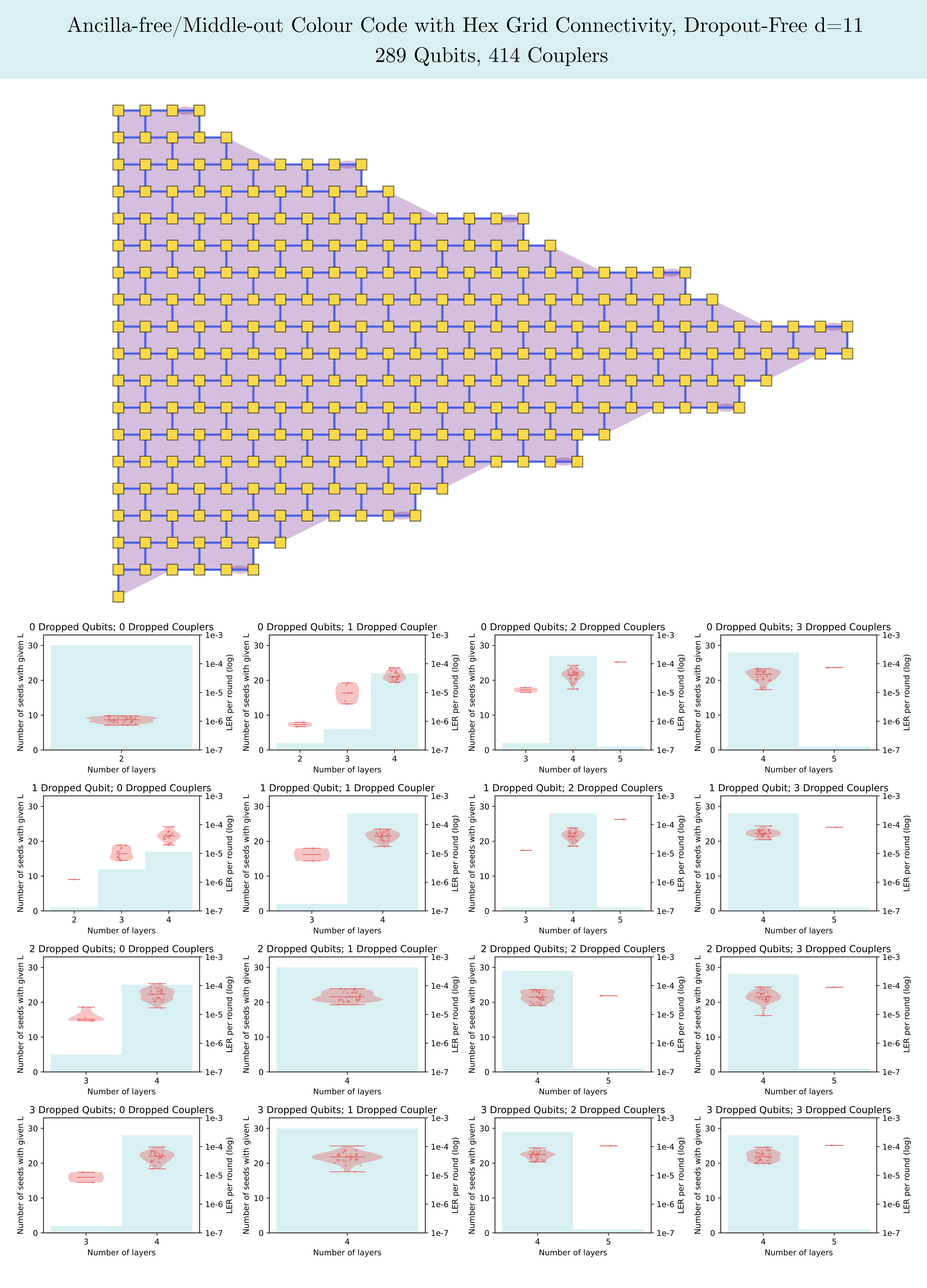}
\end{figure}
\begin{figure}
    \centering
    \includegraphics[width=1\linewidth]{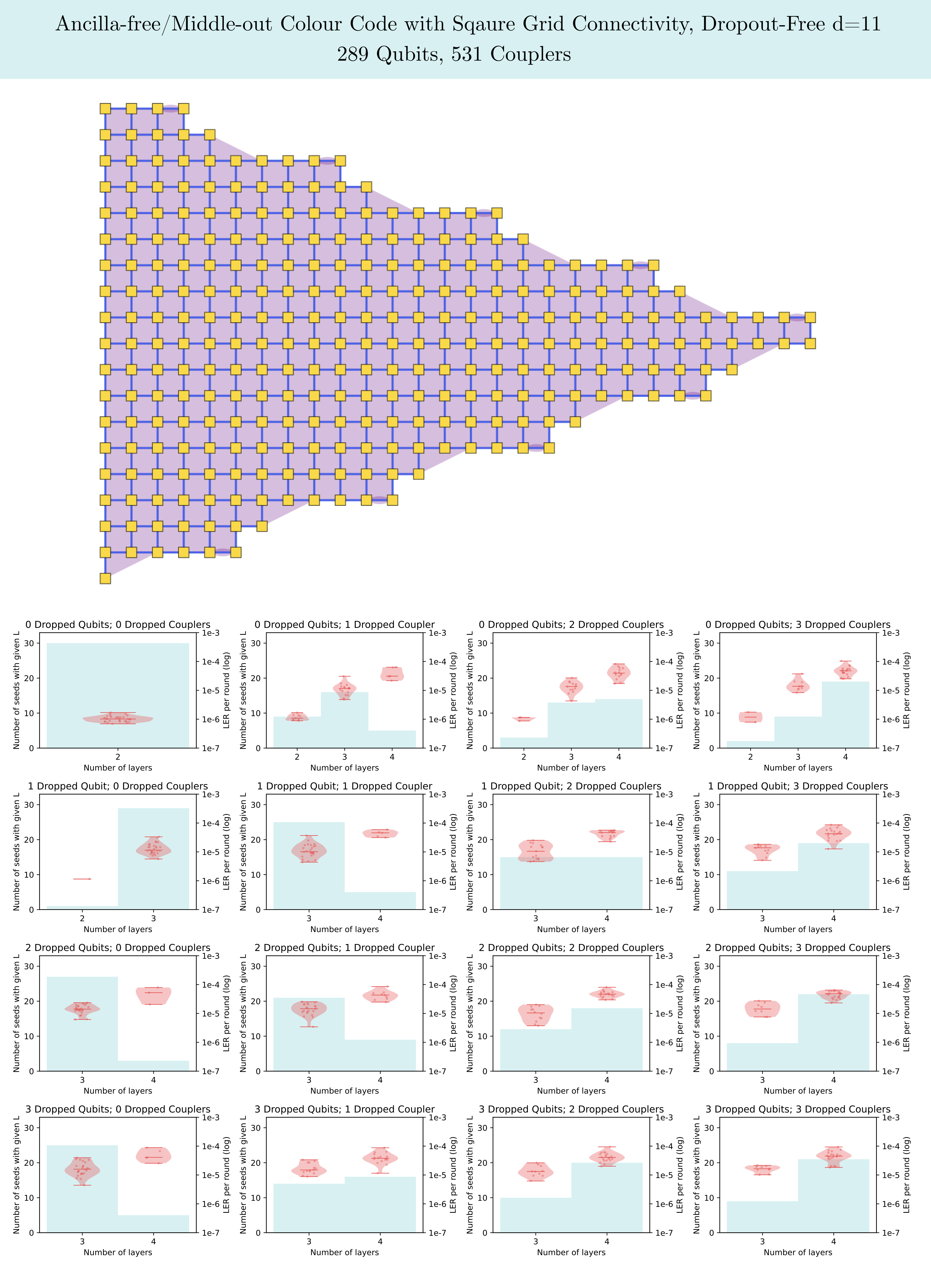}
\end{figure}
\begin{figure}
    \centering
    \includegraphics[width=1\linewidth]{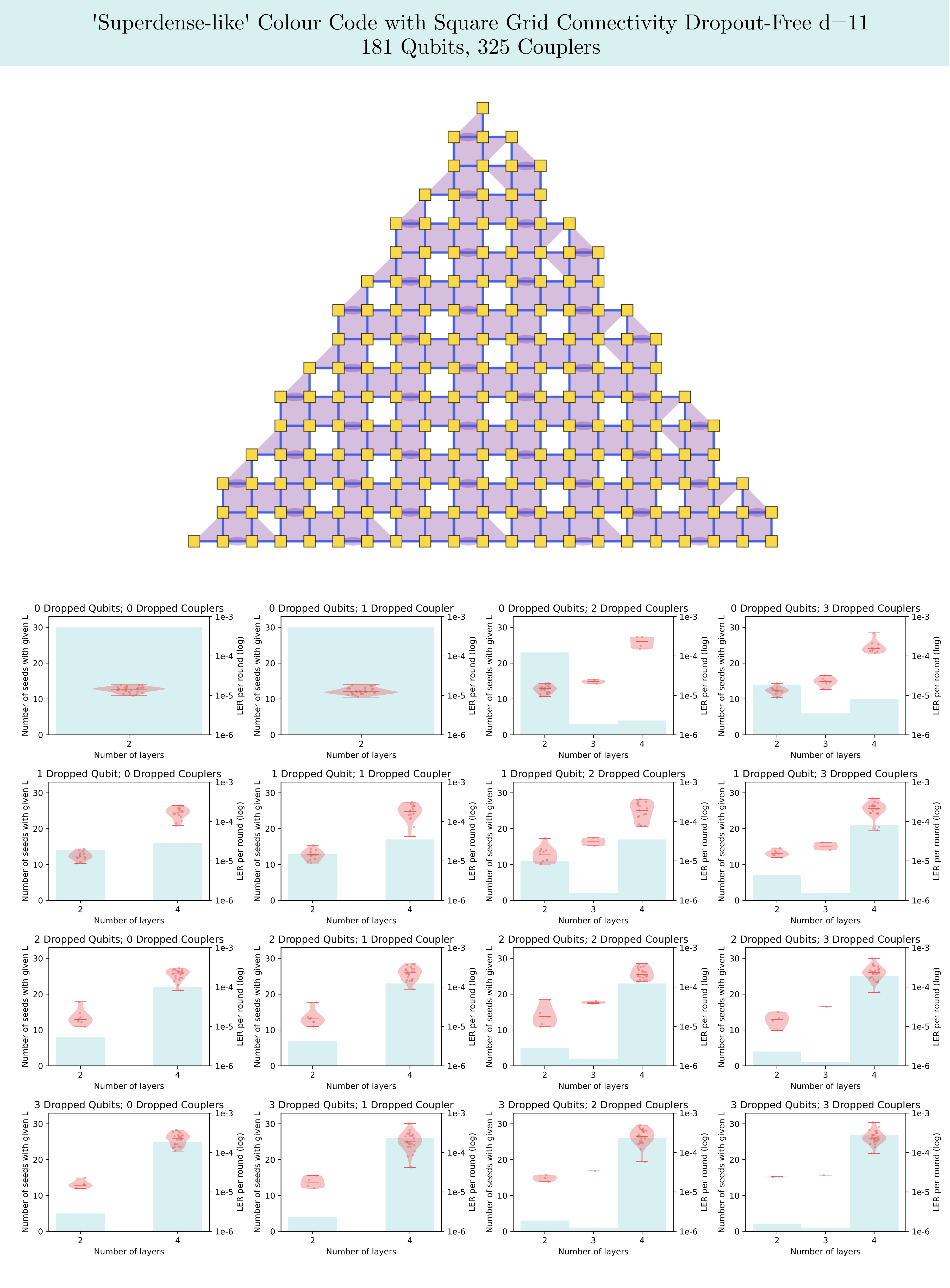}
\end{figure}

\end{appendix}

\end{document}